\documentclass[12pt]{article}

\usepackage{parskip} 
\usepackage{mathtools}
\usepackage{amsfonts}
\usepackage{amssymb}
\usepackage{amsthm}
\usepackage{amsmath}
\usepackage[utf8]{inputenc} 
\usepackage[svgnames]{xcolor}
\usepackage[colorlinks]{hyperref} 
\usepackage{tikz}       
\usepackage{fancyhdr}   
\usepackage{titlesec}   
\usepackage{times}
\usepackage{changepage}
\usepackage{graphicx}
\usepackage{float}
\usepackage{bbm}        
\usepackage{pdfpages}
\usepackage{enumitem}   
\usepackage{caption}    
\usepackage{subcaption} 
\usepackage{courier}    
\usepackage{adjustbox}  
\usepackage{booktabs}
\usepackage{multirow}
\usepackage{colortbl}   
\usepackage{authblk}
\usepackage[round]{natbib}

\titlelabel{\thetitle.\;}             

\newtheorem{algorithm}{Algorithm}

\hypersetup{
	citecolor  = {gray},
	colorlinks = true,
	linkcolor  = blue
}        

\hypersetup{pdfborder={0 0 0}}  
\theoremstyle{plain}

\newcommand*\dif{\mathop{}\!\mathrm{d}}

\DeclareMathOperator*{\argmax}{\arg\max}
\title{
	\bf Estimation and Specification Test for Diffusion Models with Stochastic Volatility
}
\author[1]{L\'opez-P\'erez, A. \thanks{\textit{Contact:} \href{mailto:alejandralopez.perez@usc.es}{alejandralopez.perez@usc.es}. The authors gratefully thank Spanish National Research Council for the computing resources of the Supercomputing Center of Galicia (CESGA). The authors acknowledge support from grant MTM2016-76969-P from the Spanish Ministry of Economy and Competitiveness.} }
\author[1]{Febrero-Bande, M.}
\author[1]{Gonz\'alez-Manteiga, W.}
\affil[1]{Department of Statistics, Mathematical Analysis and Optimization. Universidade de Santiago de Compostela}
\date{}

\begin{document}
\maketitle

\begin{center}
	\begin{minipage}{0.9\textwidth}
		\vskip 0.5cm
		\small 
		\textbf{Abstract.} Given the importance of continuous--time stochastic volatility models to describe the dynamics of interest rates, we propose a goodness--of--fit test for the parametric form of the drift and diffusion functions, based on a marked empirical process of the residuals. The test statistics are constructed using a continuous functional (Kolmogorov--Smirnov and Cramér--von Mises) over the empirical processes. In order to evaluate the proposed tests, we implement a simulation study, where a bootstrap method is considered for the calibration of the tests. As the estimation of diffusion models with stochastic volatility based on discretely sampled data has proven difficult, we address this issue by means of a Monte Carlo study for different estimation procedures. Finally, an application of the procedures to real data is provided.

	\end{minipage}
\end{center}

\vspace{0.5cm}


\section{Introduction} 

Over the last five decades, continuous-time models have proven to be an essential part of the financial econometrics field. A large body of literature for the term structure of interest rates is written in continuous-time \citep{merton1975asymptotic} since the seminal works of \cite{merton1973theory} and \cite{black1973pricing}. Different specifications have been proposed, such as the time-homogeneous diffusion given by a stochastic differential equation (SDE) driven by a Wiener process $W_t$,
\begin{equation} \label{eq:SDEuni}
\dif r_t = m(r_t, \boldsymbol{\theta})\dif t + \nu(r_t, \boldsymbol{\theta}) \dif W_t,
\end{equation}
defined on a complete probability space $\big(\Omega, \lbrace \mathcal{F}_t \rbrace_{t \in [0,T]}, \mathbb{P}\big)$, where $\Omega$ is a nonempty set, $\mathcal{F}$ is a $\sigma$-algebra of subsets of $\Omega$ and $\mathbb{P}$ is a probability measure, $\mathbb{P}(\Omega) = 1$. The process $r_t \in \mathbb{R}$ evolves over the interval $[0,T]$ in continuous time, according to the drift $m(\cdot)$ and diffusion $\nu(\cdot)$ functions. We work under a parametric framework, where $\boldsymbol{\theta}$ is an unknown parameter vector such that $\boldsymbol{\theta} \in \Theta \subset \mathbb{R}^d$ with $d$ a positive integer and $\Theta$ a compact set, and ${m(\cdot, \boldsymbol{\theta})\colon \mathbb{R} \times \Theta \rightarrow \mathbb{R}}$ and ${\nu(\cdot, \boldsymbol{\theta}) \colon \mathbb{R} \times \Theta \rightarrow (0,\infty)}$.

In order to determine if the model is appropriated for a given time series, the parametric form of both drift and volatility functions can be tested. There exist several proposals for continuous-time model specification, such as \cite{ait1996testing}, \cite{gao2004adaptive}, \cite{hong2004nonparametric}, \cite{chen2008test}, who used the marginal density function of the process; \cite{dette2003test} and \cite{dette2006estimation}, used a test statistic based on the $L^2$-distance between the diffusion function under the null hypothesis and the alternative; \cite{arapis2006empirical}, \cite{li2007testing}, \cite{gao2008specification} and \cite{chen2019nonparametric} proposals were based on smoothing techniques; \cite{fan2001generalized} and \cite{fan2003reexamination} developed a test based on a likelihood ratio test; \cite{dette2008testing} and \cite{podolskij2008range} proposals were based on stochastic processes of the integrated volatility; and \cite{monsalve2011goodness} and \cite{chen2015asymptotically} tests were based on empirical regression processes.

The empirical evidence obtained from the goodness-of-fit tests for the one-factor model in~\eqref{eq:SDEuni} proved unsatisfactory empirical fits and suggested that more flexible specifications for the volatility function were needed to capture the dynamics of returns of interest rates. Therefore, the literature has moved towards two-factor formulations, allowing the volatility function to incorporate a source of random variation, leading to a continuous-time stochastic volatility (SV) model, such as
\begin{align}
\label{eq:SVmodel1}
\dif r_t &= m_1(r_t, \boldsymbol{\theta}) \dif t + \sigma_t \nu_1(r_t, \boldsymbol{\theta}) \dif W_{1,t}, \\
\label{eq:SVmodel2}
\dif g(\sigma_t) &= m_2(g(\sigma_t), \boldsymbol{\theta}) \dif t + \nu_2(g(\sigma_t), \boldsymbol{\theta}) \dif W_{2,t},
\end{align}
where the functions $g, m_1, \nu_1, m_2$ and $\nu_2$ are sufficiently smooth and satisfy growth conditions to obtain existence and uniqueness for the stochastic differential equation solution, and $\boldsymbol{\theta} \in \Theta \subset \mathbb{R}^d$ is an unknown parameter vector. Several parametrizations of~\eqref{eq:SVmodel1}--\eqref{eq:SVmodel2} have been examined, see \cite{hull1987pricing}, \cite{heston1993closed}, \cite{andersen1997estimating}, \cite{gallant1998reprojecting}, \cite{eraker2001mcmc}, or \cite{christoffersen2009shape}, among others. 

Given the importance of the volatility in the financial market --a measure of risk that impacts in portfolio selection, option pricing, risk management or hedging--, its misspecification could lead to serious consequences. To overcome this, goodness-of-fit test should be used to study the adequacy of the proposed model towards the dynamics of the volatility. Some recent literature have addressed this issue, \citeauthor{lin2013goodness} (\citeyear{lin2013goodness}, \citeyear{lin2016goodness}) proposed a test based on the deviations between the empirical characteristic function and the parametric counterpart; \cite{lin2014bickel} considered a Bickel-Rosenblatt type test; \cite{zu2015nonparametric} used the $L^2$-distance to measure the discrepancy between the kernel and parametric deconvolution density estimator of an integrated volatility density; \cite{vetter2015estimation} test was based on a  Kolmogorov-Smirnov statistic; \cite{bull2017semimartingale} proposed a wavelet-based test; \cite{ebner2018fourier} used Fourier methods; \cite{christensen2019realized} proposal was based on the empirical distribution function; and \cite{li2021volatility} proposed a test for the integrated volatility of volatility.

In the present paper, we propose goodness-of-fit tests based on empirical processes, extending the methodology proposed by  \cite{monsalve2011goodness} to diffusion models with stochastic volatility, following the ideas suggested in \cite{gonzalez2017goodness}. Many goodness-of-fit test in the literature for continuous-time models, such as~\eqref{eq:SVmodel1}--\eqref{eq:SVmodel2}, focus on the stationary distribution of the volatility $\sigma_t^2$, however, we test that the diffusion and drift functions belong to a certain parametric family, that is,
\begin{align*}
\mathcal{H}_{0\nu} &\colon \nu_1 \in \lbrace \nu_1(\cdot, \boldsymbol{\theta})\colon \boldsymbol{\theta} \in \Theta \rbrace, \\
\mathcal{H}_{0m}   &\colon m_1 \in \lbrace m_1(\cdot, \boldsymbol{\theta})\colon \boldsymbol{\theta} \in \Theta \rbrace, 
\end{align*}
respectively. To construct the test statistic we use integrated regression models, an approach discussed in~\cite{stute1997nonparametric}, where the study of a marked empirical process based on residuals was introduced and, subsequently, extended to time series in~\cite{koul1999nonparametric}. The empirical regression processes-based goodness-of-fit tests have been studied by other authors, see \cite{diebolt1995nonparametric} for a nonlinear parametric regression function or \citeauthor{diebolt1999goodness} (\citeyear{diebolt1999goodness}, \citeyear{diebolt2001testing}), for an extension to nonlinear and heteroscedastic regression.

Notwithstanding the importance of goodness-of-fit tools for continuous-time models, the latent factor in the stochastic volatility model challenges its implementation. In addition to the unobserved volatility, the process is specified in continuous-time but the observations occur at discrete time points. Therefore, the estimation problem should be addressed, as it hinders the goodness-of-fit procedures. We attempt to discuss the intricacies of different implementations, though a comprehensive survey of estimation methods for continuous-time unobserved state-variable models is beyond the scope of this paper. Several methods have been proposed for the estimation of stochastic volatility models (see \citealp{chen2003bayesian}, for a review). One of the earliest proposals was quasi-maximum likelihood, introduced in \cite{harvey1994multivariate} (see, e.g., \citealp{ruiz1994quasi}; \citealp{hurn2013quasi}). Another analytical methods include the \cite{kalman1960new} filter (see \citealp{broto2004estimation} for a survey on likelihood-based methods); the generalized method of moments (see, e.g., \citealp{melino1990pricing}; \citealp{andersen1996gmm}; \citealp{sapp2009estimating}); approximate likelihood methods based on the characteristic function of the transition density \citep{bates2006maximum}; closed-form moment-based procedure \citep{dufour2009exact}; maximum likelihood using closed-form approximations and latent factor filtering \citep{ait2020maximum}. 
Simulation-based methods, though more computationally demanding, are increasingly used in the financial context. These methods include Markov Chain Monte Carlo techniques (see, e.g., \citealp{jacquier1994bayesian}; \citealp{shephard1997likelihood}; \citealp{kim1998stochastic}; \citealp{eraker2001mcmc}; \citealp{chib2002markov}; \citealp{johannes2010mcmc}; \citealp{kastner2014ancillarity}); particle filters (see, e.g., \citealp{kotecha2003gaussian}; \citealp{carvalho2010particle}; \citealp{lopes2011particle}; \citealp{kantas2015particle}); Expectation-Maximization algorithms (see, e.g., \citealp{dempster1977maximum}; \citealp{little2019statistical}); simulated maximum-likelihood (see, e.g., \citealp{danielsson1993accelerated}; \citealp{sandmann1998estimation}; \citealp{durham2006monte}); integrated nested Laplace approximations (INLA) methods \citep{rue2009approximate}.

The rest of the paper is structured as follows. Section~\ref{sec:2} provides an outline of estimation methods and Monte Carlo experiments are designed in order to discuss the finite sample performance of the procedures. In Section~\ref{sec:3}, the goodness-of-fit tests for the drift and diffusion functions are introduced, while in Section~\ref{sec:4} a simulation study of the proposed tests is implemented. Real data application to interest rate series is presented in Section~\ref{sec:5} and conclusions are drawn in Section~\ref{sec:6}.

\section{Estimation of diffusion models with latent variables} \label{sec:2}

Given a two-factor continuous-time diffusion model as in~\eqref{eq:SVmodel1}--\eqref{eq:SVmodel2}, where the state $r_t$ is observable but the volatility $\sigma_t$ is a latent factor, although the model is formulated in continuous-time, data are sampled in discrete time points. Therefore, a discretized version of the diffusion equations should be consider to estimate the model parameters. Taking a popular specification, such as
\begin{align} 
\label{eq:OU_SV_cont1}
\dif r_t &= (\alpha - \beta r_t)\dif t + \sigma_t \dif W_{1,t}, \\
\label{eq:OU_SV_cont2}
\dif \log \sigma_t^2 &= (\theta_0 - \theta_1 \log \sigma_t^2)\dif t + \xi \dif W_{2,t},
\end{align}
where volatility follows an Ornstein-Uhlenbeck process \citep{stein1991stock}, we assume that the process $\lbrace r_{t_i} \rbrace_{i=0}^n$ is observed at equispaced discrete time points $0 = t_0 < t_1 < \dots < t_n = T$ in the interval $[0,T=n\Delta]$, where the time step $\Delta$ between consecutive observations is fixed. The Euler-Maruyama \citep{maruyama1955continuous} method is commonly used as an approximation scheme, thereby, considering the SDE in~\eqref{eq:OU_SV_cont1}--\eqref{eq:OU_SV_cont2}, its discretized counterpart is given by
\begin{align} 
\label{eq:OU_SV_disc1}
r_{t_i} - r_{t_{i-1}} &= \alpha\Delta - \beta\Delta r_{t_{i-1}} + \sigma_{t_i} \sqrt{\Delta} \,\varepsilon_{1,t_i}, 
&&\varepsilon_{1,t_i} \sim N(0,1), \\ 
\label{eq:OU_SV_disc2}
\log \sigma_{t_i}^2 - \log \sigma_{t_{i-1}}^2 &= \theta_0\Delta - \theta_1\Delta \log \sigma_{t_{i-1}}^2 + \xi \sqrt{\Delta} \,\varepsilon_{2,t_i}, 
&&\varepsilon_{2,t_i} \sim N(0,1), 
\end{align}
with $i=0,1,\dots,n-1$, $t_i = i\Delta$, $r_{t_0} = r_0 \in \mathbb{R}$, and where $\varepsilon_{1,t_i}$ and $\varepsilon_{2,t_i}$ are independent identically distributed (i.i.d.) standard Gaussian variables, given the independent and Gaussian increments property of the Wiener process.

In the remainder of this section, we give an outline of estimation methods procedures to estimate the unknown parameter vector. The state-space models allow an easily interpretable and flexible framework for stochastic volatility models, therefore, we focus on estimation procedures that are maximum likelihood-based on a state space model representation. We begin introducing a filtering algorithm and, as several Monte Carlo-based approximations for state-space models are available, we also consider this Bayesian approach, as well as sequential Monte Carlo methods or particle filters.

\subsection{Kalman Filter} \label{sec:2KF}

The dynamic linear model \citep{kalman1960new} considers that the observation vector $\boldsymbol{y}_t$ is generated by the state-space model
\begin{alignat}{2}
\boldsymbol{x}_t &= \mathbf{\Phi} \boldsymbol{x}_{t-1} + \boldsymbol{w}_t,  \qquad
&&\boldsymbol{w}_t \sim \text{iid}~N(0, \mathbf{Q}), \label{eq:stateKF} \\  
\boldsymbol{y}_t &= \mathbf{A}_t \boldsymbol{x}_{t} + \boldsymbol{v}_t,  \qquad
&&\;\boldsymbol{v}_t \sim \text{iid}~N(0, \mathbf{R}),  \label{eq:spaceKF}
\end{alignat}
where $\boldsymbol{x}_t \in \mathbb{R}^p$ is the unknown state vector, $\boldsymbol{y}_t \in \mathbb{R}^q$ is the observed data vector, $\mathbf{A}_t \in \mathbb{R}^{q \times p}$ is the observation matrix, $\mathbf{\Phi} \in \mathbb{R}^{p \times p}$ is the transition matrix and we assume that $\left\lbrace \boldsymbol{w}_t \right\rbrace$ and $\left\lbrace \boldsymbol{v}_t \right\rbrace$ are uncorrelated. Equation~\eqref{eq:stateKF} is known as the \textit{state equation} and equation~\eqref{eq:spaceKF} as \textit{observation equation}. The state vector $\boldsymbol{x}_t$ is latent, which provides an adequate framework for the stochastic volatility model. Let $\boldsymbol{x}_{t\mid {t-1}}=\mathbb{E}\left[\boldsymbol{x}_t \mid Y_{t-1}\right]$ and $\mathbf{P}_{t\mid {t-1}} = \mathbb{E} \left[ (\boldsymbol{x}_t - \boldsymbol{x}_{t\mid {t-1}}) (\boldsymbol{x}_t - \boldsymbol{x}_{t\mid {t-1}})'\mid Y_{t-1}\right]$, with initial state $\boldsymbol{x}_0 \sim N(\boldsymbol{x}_{0\mid 0}, \mathbf{P}_{0\mid 0})$, given the data $Y_s = \lbrace y_1, \dots, y_s \rbrace$ we have 
\begin{alignat}{2}
\label{eq:den_prop}
\boldsymbol{x}_{t} &\mid Y_{t-1} &&\sim N(\boldsymbol{x}_{t\mid{t-1}}, \mathbf{P}_{t\mid{t-1}}), \\ 
\label{eq:den_pred}
\boldsymbol{y}_{t} &\mid Y_{t-1} &&\sim N(\boldsymbol{y}_{t\mid{t-1}}, \mathbf{\Sigma}_{t\mid{t-1}}), \\ 
\label{eq:den_filt}
\boldsymbol{x}_{t} &\mid Y_{t}   &&\sim N(\boldsymbol{x}_{t\mid t}, \mathbf{P}_{t\mid t}),
\end{alignat}
which are the \textit{propagation}, \textit{predictive} and \textit{filtering density}, respectively, with 
\begin{alignat*}{2}
\boldsymbol{x}_{t\mid{t-1}}&=\mathbf{\Phi} \boldsymbol{x}_{{t-1}\mid{t-1}}, \qquad 
&&\mathbf{P}_{t\mid{t-1}} = \mathbf{\Phi} \mathbf{P}_{{t-1}\mid{t-1}} \mathbf{\Phi}' + \mathbf{Q}, \\
\boldsymbol{y}_{t\mid{t-1}}&=\mathbf{A}_t \boldsymbol{x}_{t\mid{t-1}}, \qquad 
&&\mathbf{\Sigma}_{t} = \mathbf{A}_t \mathbf{P}_{t\mid{t-1}} \mathbf{A}_t' + \mathbf{R}, \\
\boldsymbol{x}_{t\mid t}   &=\boldsymbol{x}_{t\mid{t-1}} + \mathbf{K}_t \boldsymbol{\varepsilon}_t, \qquad 
&&\mathbf{P}_{t\mid t} = \mathbf{P}_{t\mid{t-1}} - \mathbf{K}_t \mathbf{\Sigma}_{t} \mathbf{K}_t',
\end{alignat*}
where $\boldsymbol{\varepsilon}_t = \boldsymbol{y}_t - \mathbf{A}_t \boldsymbol{x}_{t\mid{t-1}}$ are the prediction errors and $\mathbf{K}_t = \mathbf{P}_{t\mid{t-1}} \mathbf{A}_t' \mathbf{\Sigma}_{t}^{-1}$ is the \textit{Kalman gain}. In state space models the aim is usually the estimation of the latent state vector $\boldsymbol{x}_t$ through filtering, for which we need to estimate the marginal distribution of the state vector given the observations, $p(\boldsymbol{x}_t \mid Y_t)$. In Equations~\eqref{eq:den_prop}--\eqref{eq:den_filt} linearity and the gaussianity of errors are assumed, but more general state space models can be considered. For this linear model, the Kalman Filter \citep{kalman1960new} can be used to estimate the distribution $p(\boldsymbol{x}_t \mid Y_t)$ and computing the likelihood using the innovations. The discretized OU model in~\eqref{eq:OU_SV_disc1}--\eqref{eq:OU_SV_disc2} is not linear so, in order to use the Kalman Filter with the state-space model~\eqref{eq:stateKF}--\eqref{eq:spaceKF}, we first need to linearize it. Taking the residuals from the linear regression, $e_{t_i} = [r_{t_i} - \alpha\Delta + (1 - \beta\Delta) r_{t_{i-1}}] / \sqrt{\Delta} = \sigma_{t_i} \varepsilon_{1,t_i} $, we define the logarithm of the squared residuals
\begin{equation*}
y_{t_i} = \log e_{t_i}^2 \quad \text{and} \quad h_{t_i} = \log \sigma_{t_i}^2,
\end{equation*}
therefore, the model can be linearized as follows, parameterizing $\phi_0 = \Delta\theta_0$, $\phi_1 = (1-\theta_1\Delta)$,
\begin{alignat}{2}
\label{eq:SVspace}
y_{t_i} &= h_{t_i} + v_{t_i}, \qquad\quad &&v_{t_i} \sim \log \chi^2_1, \\
\label{eq:SVstate}
h_{t_i} &= \phi_0 + \phi_1 h_{t_{i-1}} + w_{t_i}, \qquad\quad &&w_{t_i} \sim N(0, \sigma_w^2),
\end{alignat}
with $w_{t_i} = \xi\sqrt{\Delta}\,\varepsilon_{2,t_i}$ a Gaussian distributed variable, as $\varepsilon_{2,t_i}$ is an standard Gaussian variable, and where $\sigma_w^2 = \Delta\xi^2$. The error term in the space equation~\eqref{eq:SVspace} follows a log chi-squared with one degree of freedom, as $v_{t_i} = \log \varepsilon_{1,t_i}^2$ and $\varepsilon_{1,t_i} \sim N(0,1)$. The density of the $\log \chi^2_1$ is
\begin{equation*}
f(x) = \frac{1}{\sqrt{2\pi}} \exp \bigg( -\frac{1}{2} \bigg[\exp (x) - x \bigg] \bigg), \qquad x \in \mathbb{R},
\end{equation*}
with mean $\mathbb{E} \left[ v_{t_i} \right] = \psi(1) - \log 2 \approx -1.2704$, where $\psi(\cdot)$ is a digamma function, and variance $\mathbb{V}\mathrm{ar}\left[ v_{t_i} \right] = \pi^2 / 2 \approx 4.9348$. As this density is skewed (see Figure~\ref{fig:mixturas}), it departs from the Gaussian assumption, thus different approaches have been proposed in the literature. \cite{shumway2000time} proposed modeling the $\log \chi^2_1$ with a mixture of two Gaussian variables, one centered at zero, such as $\eta_{t_i} = I_{t_i} z_{0,t_i} + (1 - I_{t_i}) z_{1,t_i}$, where $z_{0,t_i} \sim N(0,\sigma_0^2)$ and $z_{1,t_i} \sim N(\mu_1,\sigma_1^2)$ and $I_{t_i}$ is an i.i.d. Bernoulli process, $\mathbb{P}\left( I_{t_i} = 0 \right) = \pi_0$ and $\mathbb{P}\left( I_{t_i} = 1 \right) = \pi_1$, with $\pi_0 + \pi_1 = 1$. Substituting the space equation~\eqref{eq:SVspace} with 
\begin{equation} \label{eq:SVspace2}
y_{t_i} = h_{t_i} + \eta_{t_i},
\end{equation}
we have the filtering equations for this model:
\begin{equation} \label{eq:KF_alg1}
\begin{aligned}
h_{t_{i+1}\mid t_i} &= \phi_0 + \phi_1 h_{t_i\mid t_{i-1}} + \sum_{j=0}^{1} \pi_{j,t_i} K_{j,t_i} \varepsilon_{j,t_i}\\
P_{t_{i+1}\mid t_i} &= \phi_1^2 P_{t_i\mid t_{i-1}} + \sigma_w^2 - \sum_{j=0}^{1} \pi_{j,t_i} K_{j,t_i}^2 \Sigma_{j,t_i}
\end{aligned}
\end{equation}
\begin{equation} \label{eq:KF_alg2}
\begin{split} 
\epsilon_{0,t_i} &= y_{t_i} - h_{t_i\mid t_{i-1}} \\
\Sigma_{0,t_i} &= P_{t_i\mid t_{i-1}} + \sigma_0^2\\
K_{0,t_i} &= \phi_1 P_{t_i\mid t_{i-1}} / \thinspace \Sigma_{0,t_i}
\end{split}
\qquad
\begin{split} 
\epsilon_{1,t_i} &= y_{t_i} - h_{t_i\mid t_{i-1}} - \mu_1 \\
\Sigma_{1,t_i} &= P_{t_i\mid t_{i-1}} + \sigma_1^2\\
K_{1,t_i} &= \phi_1 P_{t_i\mid t_{i-1}} / \thinspace \Sigma_{1,t_i}
\end{split}
\end{equation}
where, given the density of $y_{t_i}$ conditional to its past values $Y_{t_1}$, $f(t_i \mid t_{i-1})$,
\begin{equation*}
\pi_{j,t_i} = \frac{\pi_{1,t_i} f_1(t_i \mid t_{i-1})}{\pi_{0,t_i} f_0 (t_i \mid t_{i-1}) +  \pi_{1,t_i} f_1 (t_i \mid t_{i-1})}
\vspace{0.5\baselineskip}
\end{equation*}
and $f_j(t_i \mid t_{i-1}) = N (x_{t_i\mid t_{i-1}}+\mu_j, \Sigma_{j,t_i})$, for $j=0,1$ and $\mu_0=0$. The distribution $\pi_{j,t_i}$, for $j=0,1$, is specified \textit{a priori}, usually uniform priors are chosen, $\pi_{0,t_i} = \pi_{1,t_i} = 1/2$.
Let $\boldsymbol{\theta} = ( \phi_0, \phi_1, \sigma_w^2, \mu_1, \sigma_0^2, \sigma_1^2 )' $ be the vector of unknown parameters, maximum likelihood can be used for estimation, maximizing the log-likelihood $\ln \thinspace \mathcal{L}_Y(\boldsymbol{\theta})$ given by
\begin{equation} \label{eq:kf_loglike}
\ln \thinspace \mathcal{L}_Y(\boldsymbol{\theta}) = \sum_{i=1}^{n} \ln \left( \sum_{j=0}^{1} \pi_{j,t_i} N \Big(x_{t_i\mid t_{i-1}}+\mu_j, \sigma_j^2 \Big) \right),
\end{equation}

Algorithms like the \textit{EM} \citep{dempster1977maximum} or the Newton-Raphson can be considered to maximize the log-likelihood in~\eqref{eq:kf_loglike}, as in \cite{shumway1982approach} and \cite{jones1980maximum}, for an example of both approaches.

\subsection{Markov Chain Monte Carlo}

Given the discrete version~\eqref{eq:SVspace}--\eqref{eq:SVstate} of the SDE in~\eqref{eq:OU_SV_cont1}--\eqref{eq:OU_SV_cont2}, we have the popular parametrization of the discretized stochastic volatility model,
\begin{alignat}{2}
\label{eq:SVdisc1}
e_{t_i} &= \exp \left( h_{t_i} / 2 \right) \epsilon_{t_i}, \qquad &&\varepsilon_{t_i} \sim N(0,1), \\
\label{eq:SVdisc2}
h_{t_i} &= \phi_0 + \phi_1 h_{t_{i-1}} + w_{t_i},           \qquad &&w_{t_i} \sim N(0, \sigma_w^2)
\end{alignat}
where $e_{t_i}$ are the residuals from the linear regression $r_{t_i} = \alpha\Delta + (1 - \beta\Delta) r_{t_{i-1}} + e_ {t_i}$, with $h_{t_i} = \log \sigma_{t_i}^2$, $\phi_0 = \Delta\theta_0$, $\phi_1 = (1-\theta_1\Delta)$ and $\sigma_w^2 = \Delta\xi^2$. The model in~\eqref{eq:SVdisc1} can be linearized by taking the logarithm of the squared observations, $y_{t_i} = \ln e_{t_i}^2$, as in~\eqref{eq:SVspace}. In the Bayesian approach of the estimation problem, the unknown parameter vector $\boldsymbol{\theta} = (\phi_0, \phi_1, \sigma_w^2)'$, with $\boldsymbol{\theta} \in \Theta \subset \mathbb{R}^3$, we consider a prior distribution of $\boldsymbol{\theta}$ over the parameter space $\Theta$. Given the prior distribution $\pi (\boldsymbol{\theta})$, we use Bayes' Theorem to obtain the posterior distribution of the parameter vector, 
\begin{equation*}
p(\boldsymbol{\theta} \mid \boldsymbol{e}) = \frac{p(\boldsymbol{e} \mid \boldsymbol{\theta}) \pi(\boldsymbol{\theta})}{\int_{\Theta} p(\boldsymbol{e} \mid \boldsymbol{\theta}) \pi (\boldsymbol{\theta}) \dif \boldsymbol{\theta}} \propto p(\boldsymbol{e} \mid \boldsymbol{\theta}) \pi (\boldsymbol{\theta}),
\end{equation*}
where $\boldsymbol{e} = \lbrace e_{t_i} \rbrace_{i=0}^{n}$ is the vector of residuals.
As a closed form solution might not exist, or its calculation is too difficult, sampling methods can be used to overcome this problem, such as Markov Chain Monte Carlo (MCMC) procedures, for example. Different MCMC procedures have been proposed to estimate the SV model (see \citealp{shephard1993fitting}; \citealp{jacquier1994bayesian}, for initial proposals), where the focus is targeted to the posterior density $p(\boldsymbol{\theta}, \boldsymbol{h} \mid \boldsymbol{e})$, with $\boldsymbol{h} = \lbrace h_{t_i} \rbrace_{i=0}^{n}$, as the direct analysis of $p(\boldsymbol{\theta} \mid \boldsymbol{e})$ is not possible and the likelihood function $\mathcal{L}(\boldsymbol{e} \mid \boldsymbol{\theta})$ is intractable. Through Bayes' Theorem we have that $p(\boldsymbol{h}, \boldsymbol{\theta} \mid \boldsymbol{e}) \propto p(\boldsymbol{e} \mid \boldsymbol{h}) p(\boldsymbol{h} \mid \boldsymbol{\theta}) p(\theta)$, and via MCMC methods we can sample from $p(\boldsymbol{\theta}, \boldsymbol{h} \mid \boldsymbol{e})$. 
Using the Gibbs sampler we can produce samples from the posterior $p\big(h_{t_i} \mid \boldsymbol{h}_{-t_i}, e_{t_i}, \boldsymbol{\theta}\big)$, for $i=1, \dots, n$ where $\boldsymbol{h}_{-t_i} = (h_{t_0}, \dots, h_{t_{i-1}}, h_{t_{i+1}}, \dots, h_{t_n})'$, and $p\big(\boldsymbol{\theta} \mid \boldsymbol{e}, \boldsymbol{h}\big)$, so that these samples will converge to those generated from $p\big(\boldsymbol{\theta}, \boldsymbol{h} \mid \boldsymbol{e}\big)$. Let $\boldsymbol{h}_{t_{a:b}} = (h_{t_a}, \dots, h_{t_b})'$, the Gibbs sampler algorithm for the discrete model~\eqref{eq:SVdisc1}--\eqref{eq:SVdisc2} is given in Algorithm~\ref{alg:gibbs}.
\vspace{0.5\baselineskip}

\begin{algorithm}[Gibbs sampling algorithm]
	\label{alg:gibbs}
	For $j = 0, \dots, n$, the Gibbs sampler proceeds as follows, setting $j=0$:
	
	\begin{enumerate}
		\item Initialize $\boldsymbol{h}^{(0)}$ and $\boldsymbol{\theta}^{(0)}$.
		\item Sample $\boldsymbol{h}^{(j+1)}$ from $p\big(\boldsymbol{h} \mid \boldsymbol{e}, \boldsymbol{\theta}^{(j)}\big)$. 
		\item Sample $\boldsymbol{\theta}^{(j+1)}$ from $p\big(\boldsymbol{\theta} \mid \boldsymbol{e}, \boldsymbol{h}^{(j+1)}\big)$.
		\begin{enumerate}
			\item Sample $\sigma_w^2 \mid \boldsymbol{e}, \boldsymbol{h}, \phi_0, \phi_1$.
			\item Sample $\phi_1 \mid \boldsymbol{h}, \phi_0, \sigma_w^2$.
			\item Sample $\phi_0 \mid \boldsymbol{h}, \phi_1, \sigma_w^2$.
		\end{enumerate}
		\item Set $j=j+1$ and go to $(2)$.
	\end{enumerate}
\end{algorithm} 
\vspace{0.5\baselineskip}

Step~(2) in Algorithm~\ref{alg:gibbs} can be implemented by using a Metropolis-Hastings algorithm, where $p\big(\boldsymbol{h} \mid \boldsymbol{e}, \boldsymbol{\theta}^{(j)}\big)$ is decomposed into conditionals, $p\big(h_{t_i} \mid \boldsymbol{h}_{-t_i}^{(j)}, \boldsymbol{e}, \boldsymbol{\theta}^{(j)}\big)$, for $i=1, \dots, n$, to sample $\boldsymbol{h}^{(j+1)}$. To sample $(h_{t_0} \mid \boldsymbol{\theta}, \sigma_w^2, h_{t_1})$, given $h_{t_0} \sim N(m_0, c_0)$ and $h_{t_1} \mid h_{t_0} \sim N(\phi_0 + \phi_1 h_{t_0}, \sigma_w^2)$, we can use Bayes' theorem leading to
\begin{equation*}
h_{t_0} \mid h_{t_1} \sim N(m_1, c_1),
\end{equation*}
where $m_1 = c_1 \left[c_0^{-1}m_0 + \phi_1 \sigma_w^{-2} (h_1 - \phi_0)\right]$ and $c_1 = (c_0^{-1} + \phi_1^2 \sigma_w^{-2})^{-1}$.
Regarding the conditional prior distribution of $h_{t_i}$, for $i = 1, \dots, n-1$ we have
\begin{equation*}
\begin{pmatrix} h_{t_i} \\ h_{t_{i+1}}\end{pmatrix} \sim N \left[ \begin{pmatrix} \phi_0 + \phi_1 h_{t_{i-1}} \\ (1+\phi_1)\phi_0 + \phi_1^2 h_{t_{i-1}}\end{pmatrix} ,
\sigma_w^2 \begin{pmatrix} 1 & \phi_1 \\ \phi_1 & 1+\phi_1^2\end{pmatrix}\right],
\end{equation*}
therefore
\begin{equation*}
\begin{aligned}
(h_{t_i} \mid h_{t_{i-1}}, h_{t_{i+1}}, \boldsymbol{\theta}, \sigma_w^2 ) &~\sim~ N(\mu_{t_i}, \nu^2),\\
(h_{t_n} \mid h_{t_{n-1}}, \boldsymbol{\theta}, \sigma_w^2)           &~\sim~ N(\mu_{t_n}, \sigma_w^2),
\end{aligned}
\end{equation*}
where
\begin{align*}
\mathbb{E}\left[ h_{t_i} \mid h_{t_{i-1}}, h_{t_{i+1}}, \boldsymbol{\theta}, \sigma_w^2 \right]                 &= \mu_{t_i} = \left(\frac{1-\phi_1}{1+\phi_1^2}\right)\phi_0 + \left(\frac{\phi_1}{1+\phi_1^2}\right) (h_{t_{i-1}}+h_{t_{i+1}}),\\[1.1ex]
\mathbb{V}\textnormal{ar} \left[ h_{t_i} \mid h_{t_{i-1}}, h_{t_{i+1}}, \boldsymbol{\theta}, \sigma_w^2 \right] &=  \nu^2 = \sigma_w^2 (1+\phi_1^2)^{-1} \\
\intertext{and}
\mathbb{E}\left[ h_{t_n} \mid h_{t_{n-1}}, \boldsymbol{\theta}, \sigma_w^2 \right] &= \mu_{t_n} = \phi_0 + \phi_1 h_{t_{n-1}}.
\end{align*}

We can sample $h_{t_i}$ via independent Metropolis-Hastings, let $f_N(t_i \mid a, b)$ denote the Gaussian density distribution with mean $a$ and variance $b$, the full conditional distribution of $h_{t_i}$ is given by
\begin{equation*}
p(h_{t_i} \mid \boldsymbol{h}_{-t_i}, \boldsymbol{e}, \boldsymbol{\theta}, \sigma_w^2) = p(h_{t_i} \mid h_{t_{i-1}}, h_{t_{i+1}}, \boldsymbol{\theta}, \sigma_w^2) \; p(\boldsymbol{e} \mid h_{t_i}) = f_N (h_{t_i}; \mu_t, \nu^2) \;  f_N(e_{t_i};0,\exp (h_{t_i}) ).
\end{equation*}

Given that
\begin{equation*}
\log p(e_{t_i} \mid h_{t_i}) = \textnormal{const} - \frac{1}{2} h_{t_i} - \frac{e_{t_i}^2}{2} \exp(-h_{t_i})
\end{equation*}
and that a Taylor expansion of $\exp(-h_{t_i})$ around $\mu_{t_i}$ leads to
\begin{align*}
\log p(e_{t_i} \mid h_{t_i}) &\approx \textnormal{const} - \frac{1}{2} h_{t_i} - \frac{e_{t_i}^2}{2} \Big[\exp (-\mu_{t_i}) - (h_{t_i} - \mu_{t_i}) \exp (-\mu_{t_i}) \Big], \\[1.1ex]
g(h_{t_i}) &= \exp \Big( -\frac{1}{2} h_{t_i} \left[1 - e_{t_i}^2 \exp (-\mu_t)\right] \Big),
\end{align*}
by combining $f_N(h_{t_i}; \mu_{t_i}, \nu^2)$ and $g(h_{t_i})$ we have the proposal distribution 
\begin{equation*}
p(h_{t_i} \mid h_{-{t_i}}, \boldsymbol{e}, \boldsymbol{\theta}, \sigma_w^2) \equiv N(h_{t_i}; \tilde{\mu}_{t_i}, \nu^2),
\end{equation*}
where $\tilde{\mu}_{t_i} = \mu_{t_i} + \frac{1}{2} \nu^2 (e_{t_i}^2 \exp (-\mu_{t_i}) - 1)$. Algorithm~\ref{alg:M-H} provides the independent Metropolis-Hastings algorithm, where the acceptance probability is given in step~(3).
\vspace{0.5\baselineskip}

\begin{algorithm}[Metropolis-Hastings algorithm]
	\label{alg:M-H}
	For $i = 1, \dots, n$ and $j=0,\dots,l$, the independent Metropolis-Hastings algorithm proceeds as follows:
	
	\begin{enumerate}
		\item Current state $h_{t_i}^{(j)}$, 
		\item Sample $h_{t_i}^*$ from $N(\tilde{\mu}_{t_i}, \nu^2)$
		\item Compute the acceptance probability
		\begin{equation*}
		\alpha = \min \Bigg\lbrace 1, \frac{f_N\big(h_{t_i}^*; \mu_{t_i}, \nu^2\big) \; f_N\big(e_{t_i}; 0, \exp (h_{t_i}^*) \big)}{f_N\big(h_{t_i}^{(j)}; \mu_{t_i}, \nu^2\big) \; f_N\big(e_{t_i};0,\exp (h_{t_i}^{(j)}) \big)} \; 
		\frac{f_N\big(h_{t_i}^{(j)}; \tilde{\mu}_{t_i}, \nu^2\big)}{f_N\big(h_{t_i}^*; \tilde{\mu}_{t_i}, \nu^2\big)\phantom{\big|}} \Bigg\rbrace
		\end{equation*}
		\item New state: \\[-1.5ex]
		\begin{align*}
		h_{t_i}^{(j+1)} = \begin{cases}
		h_{t_i}^*  \; &\textnormal{w.p.} \quad \alpha  \\[1.05ex]
		h_{t_i}^{(j)} \; &\textnormal{w.p.} \quad 1 - \alpha
		\end{cases}
		\end{align*}
	\end{enumerate}
\end{algorithm} 
\vspace{0.5\baselineskip}

Regarding the sampling of the hyperparameters $\boldsymbol{\theta} = (\phi_0, \phi_1, \sigma_w^2)'$ --step~(3) in Algorithm~\ref{alg:gibbs}--, setting the initial log volatility $h_{t_0} \sim N(m_0, c_0)$ and $\boldsymbol{\phi} = (\phi_0, \phi_1)'$, the prior distributions of $\boldsymbol{\phi}$ and $\sigma_w^2$ are
\begin{equation*}
\begin{aligned}
\boldsymbol{\phi} \mid \sigma_w^2 &~\sim~ N(\boldsymbol{\theta}^{(0)}, \sigma_w^2 \mathbf{V}_0), \\[1.1ex]
\sigma_w^2                        &~\sim~ IG\left(\frac{\nu_0}{2}, \frac{\nu_0 s_0^2}{2}\right),
\end{aligned}
\end{equation*}
respectively. Conditional on $\boldsymbol{h}_{t_{0:n}}$, the posterior distribution of $\boldsymbol{\phi}$ and $\sigma_w^2$ is
\begin{equation*}
\begin{aligned}
(\boldsymbol{\phi} \mid \sigma_w^2, \boldsymbol{e}, h_{t_{0:n}}) &~\sim~ N(\boldsymbol{\phi}^{(1)}, \sigma_w^2 \mathbf{V}_1), \\
(\sigma_w^2  \mid \boldsymbol{e}, \boldsymbol{h}_{t_{0:n}})      &~\sim~ IG\left(\frac{\nu_1}{2}, \frac{\nu_1 s_1^2}{2}\right),
\end{aligned}
\end{equation*}
given that $\nu_1 = \nu_0 + n$, $$\mathbf{X} = \begin{pmatrix} 1 & h_{t_0} \\ \vdots & \vdots \\ 1 & h_{t_{n-1}} \end{pmatrix}$$ and
\begin{equation*}
\begin{aligned}
\mathbf{V}_1            &= (\mathbf{V}_0^{-1} + \mathbf{X}'\mathbf{X})^{-1}, \\
\boldsymbol{\phi}^{(1)} &= \mathbf{V}_1 (\mathbf{V}_0^{-1} \boldsymbol{\phi}^{(0)} + \mathbf{X}'\boldsymbol{h}_{t_{1:n}}), \\
\nu_1 s_1^2             &= \nu_0 s_0^2 + (\boldsymbol{e} - \mathbf{X}\boldsymbol{\phi}^{(1)})' (\boldsymbol{e} - \mathbf{X}\boldsymbol{\phi}^{(1)}) + (\boldsymbol{\phi}^{(1)} - \boldsymbol{\phi}^{(0)})' \mathbf{V}_0^{-1} (\boldsymbol{\phi}^{(1)} - \boldsymbol{\phi}^{(0)}).
\end{aligned}
\end{equation*}

\subsection{Particle Filter}

Particle filters incorporate the sequential estimation approach of the Kalman Filter algorithms and the flexibility for modeling of MCMC sampling algorithms. Replacing the Kalman Filter recursions in~\eqref{eq:den_prop} and~\eqref{eq:den_filt} by
\begin{align}
\label{eq:PLdist1}
p\big(\boldsymbol{x}_{t_i} \mid Y_{t_{i-1}}\big) &= \int p\big(\boldsymbol{x}_{t_i} \mid \boldsymbol{x}_{t_{i-1}}\big) 
	p\big(\boldsymbol{x}_{t_{i-1}} \mid Y_{t_{i-1}}\big) \dif \boldsymbol{x}_{t_{i-1}}, \\
\label{eq:PLdist2}
p\big(\boldsymbol{x}_{t_i} \mid Y_{t_{i}}\big)   &= \frac{p\big(Y_{t_i} \mid \boldsymbol{x}_{t_{i}}\big) p\big(\boldsymbol{x}_{t_{i}} \mid Y_{t_{i-1}}\big)}{p\big(Y_{t_i} \mid Y_{t_{i-1}}\big)},
\end{align}
respectively, leads to a more general dynamic model, where assumptions like normality and linearity can be relaxed. However, both distributions in~\eqref{eq:PLdist1} and~\eqref{eq:PLdist2} are intractable and computationally costly. Particle filters algorithms approximate $p\big(\boldsymbol{x}_{t_{i}} \mid Y_{t_i}\big)$ by drawing a set of $l$ i.i.d. particles $\lbrace \boldsymbol{x}_{t_i}^{(j)} \rbrace_{j=1}^l$, starting with a set of i.i.d. particles $\lbrace \boldsymbol{x}_{t_{i-1}}^{(j)} \rbrace_{j=1}^l$ approximating $p \big( \boldsymbol{x}_{t_{i-1}} \mid Y_{t_{i-1}} \big)$. Since the early sequential Monte Carlo algorithm proposed by \cite{west1992modelling}, several filters have been proposed in the literature, like the Bootstrap filter or sequential importance sampling with resampling (SISR) by \cite{gordon1993novel} and the auxiliary particle filter or auxiliary SIR (ASIR) proposed by \cite{pitt1999filtering}, among others. 
\cite{liu2001combined} proposed a filter for sequential learning, a variant of the auxiliary particle filtering (APF) algorithm, that combines the APF together with a kernel approximation to $p(\boldsymbol{\theta} \mid Y_{t_{i-1}})$ using a mixture of multivariate Gaussian distributions and shrinkage parameter to provide artificial evolution for the parameter vector $\boldsymbol{\theta}$. Therefore, the posterior distribution for $\boldsymbol{\theta}$ is approximated by the normal mixture  
\begin{equation*}
p\big(\boldsymbol{\theta} \mid Y_{t_i}\big) = \sum_{j=1}^{l} N \big(m^{(j)}, h^2 V_{t_i}\big),
\end{equation*}
where $m^{(j)} = a \boldsymbol{\theta}_{t_i}^{(j)} + (1-a) \tilde{\boldsymbol{\theta}}_{t_i}$, $a=\sqrt{1-h^2}$, $\tilde{\boldsymbol{\theta}}_{t_i} = \sum_{j=1}^{l} \boldsymbol{\theta}_{t_i}^{(j)} / l$ and $V_{t_i} = \sum_{j=1}^{l} (\boldsymbol{\theta}_{t_i}^{(j)} - \tilde{\boldsymbol{\theta}}_{t_i})(\boldsymbol{\theta}_{t_i}^{(j)} - \tilde{\boldsymbol{\theta}}_{t_i})' / l$. The constant $a$ measures the extent of the shrinkage and $h$ controls the degree of overdispersion of the mixture (the choice of both parameters is discussed in \citealp{liu2001combined}). The general algorithm is displayed in Algorithm~\ref{alg:PF}.
\vspace{0.5\baselineskip}

\begin{algorithm}[Liu and West filter]
	\label{alg:PF}
	For $i = 1, \dots, n$, a general \cite{liu2001combined} filter algorithm runs as follows:	
	\begin{enumerate}
		\item Set the prior point estimates $\lbrace(\hat{\mu}_{t_{i+1}}, m_{t_i})^{(j)}\rbrace_{j=1}^l$ of $(x_{t_i}, \boldsymbol{\theta})$ where $\hat{\mu}_{t_i}^{(j)} = \mathbb{E} \left[ x_{t_{i+1}} \mid x_{t_i}^{(j)}, \boldsymbol{\theta}^{(j)} \right]$.
		\item Resample $\lbrace (\tilde{x}_{t_i}, \tilde{\boldsymbol{\theta}}_{t_i})^{(j)} \rbrace_{j=1}^{l}$ from $\lbrace (x_{t_i}, \boldsymbol{\theta}_{t_i})^{(j)} \rbrace_{j=1}^{l}$ with weights
		\begin{equation*}
			w_{t_{i+1}}^{(j)} \propto p \big( y_{t_{i+1}} \mid \hat{\mu}_{t_{i+1}}^{(j)}, m^{(j)} \big).
		\end{equation*}
		\item Propagate
			\begin{enumerate}
				\item $\lbrace \tilde{\boldsymbol{\theta}}_{t_i}^{(j)} \rbrace_{j=1}^l$ via $N \big(\tilde{m}^{(j)}, h^2 V_{t_i}\big)$,
				\item $\lbrace \tilde{x}_{t_i}^{(j)} \rbrace_{j=1}^l$ via $p \big( x_{t_{i+1}} \mid \tilde{x}_{t_i}^{(j)}, \tilde{\boldsymbol{\theta}}_{t_{i+1}}^{(j)} \big)$.
			\end{enumerate}
		\item Resample $\lbrace \big( x_{t_{i+1}}, \boldsymbol{\theta}_{t_{i+1}} \big)^{(j)} \rbrace_{j=1}^l$ from $\lbrace \big( \tilde{x}_{t_{i+1}}, \tilde{\boldsymbol{\theta}}_{t_{i+1}}^{(j)} \big)^{(j)} \rbrace_{j=1}^l$ with weights
		\begin{equation*}
			w_{t_{i+1}}^{(j)} \propto \frac{p\big( y_{t_{i+1}} \mid \tilde{x}_{t_{i+1}}^{(j)}, \tilde{\boldsymbol{\theta}}_{t_{i+1}}^{(j)} \big)}{p\big( y_{t_{i+1}} \mid \hat{\mu}_{t_{i+1}}^{(j)}, \tilde{m}^{(j)} \big) }.
		\end{equation*}
	\end{enumerate}
\end{algorithm}

\subsection{Comparative study}

This section compares the results of the different estimation procedures applied to three parametrizations of the continuous-time two-factor model with stochastic volatility: we consider a simpler model, such as~\eqref{eq:OU_SV_cont1}--\eqref{eq:OU_SV_cont2}, which does not include a level parameter; a model with a more intricate volatility function, with level parameter, with and without correlated errors. We compare the parameter estimates obtained with the different procedures under Monte Carlo settings to examine their finite sample performance. 
Through all the models and procedures considered, we first estimate the parameters of the drift function $m_1(\cdot)$ in~\eqref{eq:SVmodel1} and, subsequently, the residuals obtained are used in the procedures to estimate the parameter vector $\boldsymbol{\theta}$.

\subsubsection{Ornstein-Uhlenbeck with stochastic volatility}

Considering the Ornstein-Uhlenbeck model with stochastic volatility introduced in~\eqref{eq:OU_SV_cont1}--\eqref{eq:OU_SV_cont2} and its discretized counterpart, \eqref{eq:OU_SV_disc1}--\eqref{eq:OU_SV_disc2}, we have the discrete two-factor model
\begin{alignat*}{2} \label{eq:SDE_OU_dis}
r_{t_{i+1}} - r_{t_i} &= (\alpha - \beta r_{t_i}) \Delta + \sigma_{t_i} \sqrt{\Delta} \varepsilon_{1,t_i}, \qquad &&\varepsilon_{1,t_i} \sim N(0,1),\\
\log \sigma_{t_{i+1}}^2 &= \phi_0 - \phi_1 \log \sigma_{t_i}^2  + w_{t_i}, \qquad &&w_{t_i} \sim N(0,\sigma_w^2),
\end{alignat*}
where $\phi_0 = \theta_0 \Delta$, $\phi_1 = 1 - \theta_1 \Delta$ and $\sigma_w^2 = \Delta \xi^2$, with $i = 0,1,\dots,n-1$ and initial condition $r_{t_0} = r_0 \in \mathbb{R}$. To estimate the vector of parameters $\boldsymbol{\theta} = (\alpha, \beta, \phi_0, \phi_1, \sigma_w^2)'$ we first obtain the residuals from the linear regression $u_{t_i} = \alpha - \beta r_{t_i} + e_{t_i}$, where $u_{t_i} = (r_{t_{i+1}} - r_{t_i}) / \Delta$. Therefore, in this first step we obtain the estimates of $\alpha$ and $\beta$ and thereafter we use the procedures to estimate the rest of the parameters using the residuals of the linear regression, $e_{t_i} = \sigma_{t_i} \Delta^{-1/2} \varepsilon_{1,t_i}$. The vector of parameter values considered for data simulation is $\boldsymbol{\theta} = (\alpha, \beta, \phi_0, \phi_1, \sigma_w^2)' = (0.01, 0.3, -0.006, 0.99, 0.0225)'$, with weekly frequency $(\Delta = 1/52)$ for sample size $n \in \lbrace 520, 1040, 2080 \rbrace$, which corresponds to $T = 10, 20 \text{ and } 40$ years, respectively. A thousand realizations of random sample paths $\lbrace r_{i\Delta}\rbrace_{i=1}^n$ were generated, where the first $1000$ observations were discarded to remove the dependence on the initial value.

Tables~\ref{tab:estT10}--\ref{tab:estT40} report the estimates for the three procedures considered: Markov Chain Monte Carlo (MCMC) method, using a Metropolis-Hastings algorithm within the Gibbs sampling; the \cite{liu2001combined} filter (Particle Filter); and the \cite{kalman1960new} Filter. Tables include the mean and variance (Var) for one thousand simulations, along with the mean squared error (MSE). The estimates of the drift parameters $\alpha$ and $\beta$ were not obtained with the procedures, as mentioned, but rather fitting a linear regression. The simulation-based techniques show low MSE with the different sample sizes considered, while the Kalman Filter, thought for the smallest sample size $n$ has higher bias and variance, it particularly decreases when the observation window $T$ is extended, achieving a MSE closer to the other methods. Whilst the MCMC and the particle filter procedures do deliver accurate estimations, their computationally demanding nature and the practical implementation, highly model-dependent, are major disadvantages. On the other hand, the flexibility of the Kalman Filter can provide a good trade-off between speed and efficiency.

\begin{table}[H]
	\centering
	\renewcommand{\arraystretch}{0.9}
	\begin{tabular}{crrrr}
		\toprule
		\multicolumn{1}{c}{\textbf{Parameter}} & \multicolumn{1}{c}{\textbf{True}} & \multicolumn{1}{c}{\textbf{Mean}} & \multicolumn{1}{c}{\textbf{Var}} & \multicolumn{1}{c}{\textbf{MSE}} \\
		\midrule
		$\alpha$     & $0.01$  & $0.0195$ & $0.6764$ & $0.6765$ \\
		$\beta$      & $0.3 $  & $0.8888$ & $0.3903$ & $0.7370$ \\
		\multicolumn{5}{c}{\textbf{MCMC}} \\
		$\phi_0$     & $-0.006$ & $-0.0139$ & $0.0019   $ & $0.0020   $ \\
		$\phi_1$     & $0.99  $ & $0.9692 $ & $0.0065   $ & $0.0069   $ \\
		$\sigma_w^2$ & $0.0225$ & $0.0231 $ & $4.457\times 10^{-5}$ & $4.490\times 10^{-5}$ \\
		\multicolumn{5}{c}{\textbf{Particle Filter}} \\
		$\phi_0$     & $-0.006$ & $-0.0224$ & $0.0011   $ & $0.0013   $ \\
		$\phi_1$     & $0.99  $ & $0.9370 $ & $0.0004   $ & $0.0032   $ \\
		$\sigma_w^2$ & $0.0225$ & $0.0376 $ & $1.394\times 10^{-4}$ & $3.663\times 10^{-4}$ \\
		\multicolumn{5}{c}{\textbf{Kalman Filter}} \\
		$\phi_0$     & $-0.006$ & $-0.0420$ & $0.0506   $ & $0.0519   $ \\
		$\phi_1$     & $0.99  $ & $0.9457 $ & $0.0467   $ & $0.0487   $ \\
		$\sigma_w^2$ & $0.0225$ & $0.0410 $ & $9.213\times 10^{-4}$ & $1.265\times 10^{-3}$ \\
		\bottomrule
	\end{tabular}%
	\caption{\label{tab:estT10} Parameter estimates for the Ornstein-Uhlenbeck process with stochastic volatility, with sample size $n=520$ which corresponds to ten years ($T=10$) of weekly data ($\Delta=1/52$).}%
\end{table}%

\begin{table}[H]
	\centering
	\renewcommand{\arraystretch}{0.9}
	\begin{tabular}{crrrr}
		\toprule
		\multicolumn{1}{c}{\textbf{Parameter}} & \multicolumn{1}{c}{\textbf{True}} & \multicolumn{1}{c}{\textbf{Mean}} & \multicolumn{1}{c}{\textbf{Var}} & \multicolumn{1}{c}{\textbf{MSE}} \\
		\midrule
		$\alpha$ & $0.01$  & $0.0089$ & $0.1617$ & $0.1617$ \\
		$\beta$  & $0.3 $  & $0.5676$ & $0.1240$ & $0.1956$ \\
		\multicolumn{5}{c}{\textbf{MCMC}} \\
		$\phi_0$     & $-0.006$ & $-0.0083$ & $7.270\times 10^{-5}$ & $7.781\times 10^{-5}$ \\
		$\phi_1$     & $0.99  $ & $0.9851 $ & $6.075\times 10^{-5}$ & $8.525\times 10^{-5}$ \\
		$\sigma_w^2$ & $0.0225$ & $0.0227 $ & $3.341\times 10^{-5}$ & $3.345\times 10^{-5}$ \\
		\multicolumn{5}{c}{\textbf{Particle Filter}} \\
		$\phi_0$     & $-0.006$ & $-0.0211$ & $6.144\times 10^{-4}$ & $8.422\times 10^{-4}$ \\
		$\phi_1$     & $0.99  $ & $0.9501 $ & $2.597\times 10^{-4}$ & $1.854\times 10^{-3}$ \\
		$\sigma_w^2$ & $0.0225$ & $0.0369 $ & $1.462\times 10^{-4}$ & $3.539\times 10^{-4}$ \\
		\multicolumn{5}{c}{\textbf{Kalman Filter}} \\
		$\phi_0$     & $-0.006$ & $-0.0168$ & $3.885\times 10^{-3}$ & $5.178\times 10^{-3}$ \\
		$\phi_1$     & $0.99  $ & $0.9585 $ & $3.985\times 10^{-2}$ & $4.181\times 10^{-2}$ \\
		$\sigma_w^2$ & $0.0225$ & $0.0350 $ & $1.902\times 10^{-4}$ & $5.336\times 10^{-4}$ \\
		\bottomrule
	\end{tabular}%
	\caption{\label{tab:estT20} Parameter estimates for the Ornstein-Uhlenbeck process with stochastic volatility, with sample size $n=1040$ which corresponds to ten years ($T=20$) of weekly data ($\Delta=1/52$).}%
\end{table}%

\begin{table}[H]
	\centering
	\renewcommand{\arraystretch}{0.9}
	\begin{tabular}{crrrr}
		\toprule
		\multicolumn{1}{c}{\textbf{Parameter}} & \multicolumn{1}{c}{\textbf{True}} & \multicolumn{1}{c}{\textbf{Mean}} & \multicolumn{1}{c}{\textbf{Var}} & \multicolumn{1}{c}{\textbf{MSE}} \\
		\midrule
		$\alpha$ & $0.01$  & $0.0192$ & $0.0508$ & $0.0509$ \\
		$\beta$  & $0.3 $  & $0.4378$ & $0.0433$ & $0.0623$ \\
		\multicolumn{5}{c}{\textbf{MCMC}} \\
		$\phi_0$     & $-0.006$ & $-0.0071$ & $2.699\times 10^{-5}$ & $2.825\times 10^{-5}$ \\
		$\phi_1$     & $0.99  $ & $0.9878 $ & $2.121\times 10^{-5}$ & $2.585\times 10^{-5}$ \\
		$\sigma_w^2$ & $0.0225$ & $0.0226 $ & $2.323\times 10^{-5}$ & $2.325\times 10^{-5}$ \\
		\multicolumn{5}{c}{\textbf{Particle Filter}} \\
		$\phi_0$     & $-0.006$ & $-0.0176$ & $3.167\times 10^{-4}$ & $4.508\times 10^{-4}$ \\
		$\phi_1$     & $0.99  $ & $0.9629 $ & $1.610\times 10^{-4}$ & $8.958\times 10^{-4}$ \\
		$\sigma_w^2$ & $0.0225$ & $0.0344 $ & $1.420\times 10^{-4}$ & $2.827\times 10^{-4}$ \\
		\multicolumn{5}{c}{\textbf{Kalman Filter}} \\
		$\phi_0$     & $-0.006$ & $-0.0087$ & $3.638\times 10^{-5}$ & $1.329\times 10^{-3}$ \\
		$\phi_1$     & $0.99  $ & $0.9874 $ & $2.627\times 10^{-5}$ & $1.992\times 10^{-3}$ \\
		$\sigma_w^2$ & $0.0225$ & $0.0340 $ & $7.659\times 10^{-5}$ & $4.200\times 10^{-4}$ \\
		\bottomrule
	\end{tabular}%
	\caption{\label{tab:estT40} Parameter estimates for the Ornstein-Uhlenbeck process with stochastic volatility, with sample size $n=2080$ which corresponds to ten years ($T=40$) of weekly data ($\Delta=1/52$).}%
\end{table}%

\subsubsection{CKLS with stochastic volatility}

In this section, a more intricate model is considered, based on the CKLS model proposed in \cite{chan1992empirical}, where stochastic volatility is incorporated to the diffusion function. The CKLS model with stochastic volatility described by the Ornstein-Uhlenbeck (OU) process is given by
\begin{equation} \label{eq:SDE_CKLS}
\begin{aligned}
\dif r_t &= (\alpha - \beta r_t) \dif t + \sigma_t r_t^\gamma \dif W_{1,t}, \\
\dif \log \sigma_t^2 &= (\theta_0 - \theta_1 \log \sigma_t^2 ) + \xi \dif W_{2,t}, \\
\end{aligned}
\end{equation}
and its discretized counterpart,
\begin{equation} \label{eq:SDE_CKLS_dis}
\begin{aligned}
r_{t_{i+1}} - r_{t_i} &= (\alpha - \beta r_t) \Delta + \sigma_t r_{t_i}^\gamma \, (W_{1,t_{i+1}} - W_{1,t_i}), \\
\log \sigma_{t_{i+1}}^2 &= \phi_0 - \phi_1 \log \sigma_{t_i}^2  + \xi \, (W_{2,t_{i+1}} - W_{2,t_i}), \\
\end{aligned}
\end{equation}
where $\phi_0 = \theta_0 \Delta$ and $\phi_1 = 1 - \theta_1 \Delta$. This process has been proposed in the literature to model the short term interest rate (see \citeauthor{andersen1997estimating}, \citeyear{andersen1997estimating}; \citeyear{andersen1997stochastic}; among others), as it represents an extension of the classical stochastic volatility model to a continuous-time setting incorporating \textit{level effect}, implying that volatility depends on the level of the interest rate and inducing conditional heteroskedasticity.

\begin{figure}[H]
	\centering
	\includegraphics[width=1\linewidth]{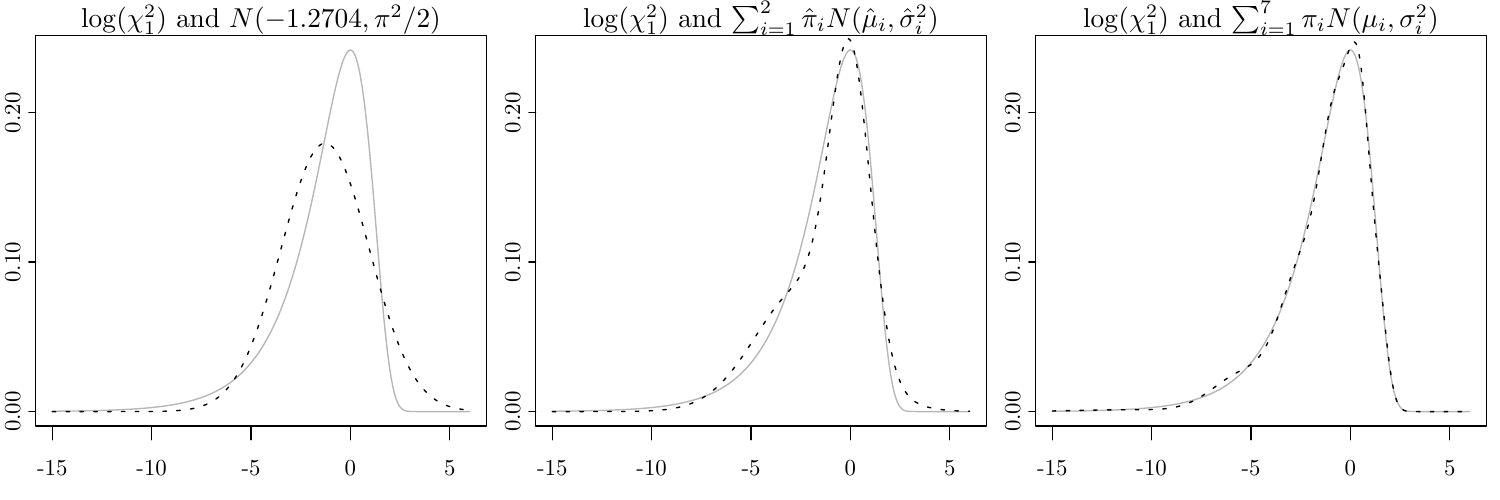}
	\caption{Comparison of the $\log \chi^2_1$ density function (solid line) and Gaussian density (left), mixture of two normal distributions (middle) and mixture of seven normal distributions (right).}
	\label{fig:mixturas}
\end{figure}

The Kalman Filter can be easily extended to allow modifications of the two-factor model. As indicated in Section~\ref{sec:2KF}, the error term in the space equation~\eqref{eq:SVspace} follows a log chi-squared with one degree of freedom, and this has motivated different approaches in the literature. \cite{shumway2000time} modeled the $\log \chi^2_1$ with a mixture of two Gaussian variables (see the filtering equations~\eqref{eq:KF_alg1}--\eqref{eq:KF_alg2}), while \cite{kim1998stochastic} proposed a seven-component Gaussian mixture (see \citealp{chib2002markov}, and \citealp{artigas2004efficient}), with weights $\pi_i$ and mean and variance ($N(\mu_i, \sigma_i^2)$ for $i\in \lbrace 1, \dots, 7 \rbrace$) given in Table~\ref{tab:mix7}. A comparison of the true $\log \chi^2_1$ distribution with a Gaussian distribution and the two and seven Gaussian mixture is illustrated in Figure~\ref{fig:mixturas}.

\begin{table}[H]
	\centering
	\renewcommand{\arraystretch}{0.9}
	\begin{tabular}{cccccccc}
		\toprule 
		Component $i$ & $1$ & $2$ & $3$ & $4$ & $5$ & $6$ & $7$ \\ 
		\midrule 
		$\pi_i$      & $0.00730$ & $0.10556$ & $0.00002$ & $0.04395$ & $0.34001$ & $0.24566$ & $0.25750$ \\
		$\mu_i$      & $-11.400$ & $-5.2432$ & $-9.8373$ & $ 1.5075$ & $-0.6510$ & $ 0.5248$ & $-2.3586$ \\
		$\sigma^2_i$ & $5.7960$  & $2.6137$  & $5.1795$  & $0.1674$  & $0.6401$  & $0.3402$  & $1.2626$  \\
		\bottomrule
	\end{tabular} 
	\caption{Components of a mixture of seven Gaussian distributions, $N(\mu_i, \sigma_i^2)$, with weights $\pi_i$.}
	\label{tab:mix7}
\end{table}

A simulation study was conducted to compare both approaches. A thousand realizations of random sample paths for the CKLS-OU model in~\eqref{eq:SDE_CKLS} were generated with weekly frequency, with $\boldsymbol{\theta} = (\alpha, \beta, \gamma, \phi_0, \phi_1, \xi)' = (0.04, 0.6, 1.5, -0.013, 0.998, 0.4)'$. Note that in the two mixture approach we estimate the parameter vector $\boldsymbol{\theta}$ and the components of the mixture --namely, $(\mu_1, \sigma_0^2, \sigma_1^2)'$, as $\mu_0 = 0$--, while in the seven mixture approach the means and variances of the Gaussian distributions remain fixed according to the values in Table~\ref{tab:mix7}. 

\begin{figure}[H]
	\centering
	\includegraphics[width=1\linewidth]{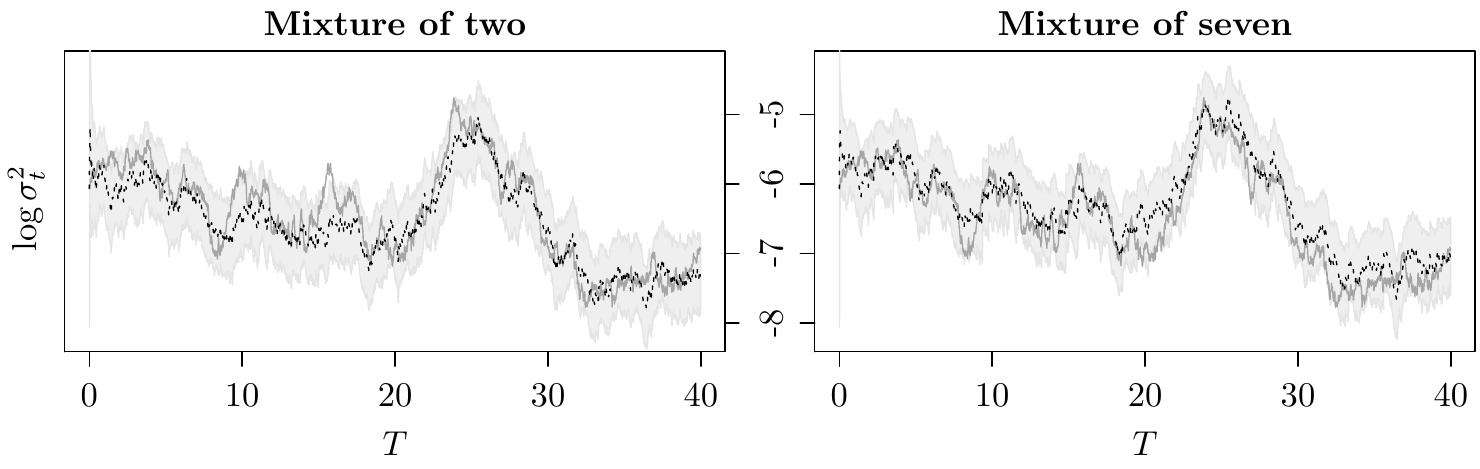}
	\caption{Comparison of the true $\log \sigma^2_t$ path (solid line) and estimated (dotted line) with a mixture of two normal distributions (left) and mixture of seven normal distributions (right). Shaded in gray is a 95\% confidence interval for both estimations.}
	\label{fig:logsig_est}
\end{figure}

The parameter estimates are summarized in Tables~\ref{tab:KF_T10}--\ref{tab:KF_T40}, for the two (Kalman Filter 2) and seven (Kalman Filter 7) mixture and three sample sizes $n \in \lbrace 520, 1040, 2080 \rbrace$. Though both methods provide a close performance, for small sample size (Table~\ref{tab:KF_T10}) the seven mixture filter provides lower mean squared error. However, larger sample sizes (Table~\ref{tab:KF_T40}) show an improvement for the two mixture filter, where the estimation of the parameter $\xi$, known as the \textit{volatility of volatility} and hard to estimate accurately, outperforms the seven mixture filter. As an example, Figure~\ref{fig:logsig_est} shows the estimated path (dotted) of the log volatility ($\log \sigma_t^2$) using the two and seven mixture approach, and a $95 \%$ confidence interval for the estimated paths (shaded).

\begin{table}[H]
	\centering
	\begin{tabular}{lrrrrrrrr}
		&       & \multicolumn{3}{c}{\textbf{Kalman Filter (2)}} &       & \multicolumn{3}{c}{\textbf{Kalman Filter (7)}} \\
		\midrule
		& \multicolumn{1}{c}{True} & \multicolumn{1}{c}{Mean} & \multicolumn{1}{c}{Var} & \multicolumn{1}{c}{MSE} &       & \multicolumn{1}{c}{Mean} & \multicolumn{1}{c}{Var} & \multicolumn{1}{c}{MSE} \\
		\midrule
		$\alpha$ & $0.04  $ & $0.0731 $ & $0.0017$ & $0.0028$ & & $0.0731 $ & $0.0017$ & $0.0028$ \\
		$\beta $ & $0.6   $ & $1.0964 $ & $0.3760$ & $0.6224$ & & $1.0964 $ & $0.3760$ & $0.6224$ \\
		$\gamma$ & $1.5   $ & $1.5305 $ & $0.0243$ & $0.0253$ & & $1.5009 $ & $0.0133$ & $0.0133$ \\
		$\phi_0$ & $-0.013$ & $-0.4977$ & $1.5604$ & $1.7948$ & & $-0.2974$ & $0.8113$ & $0.8919$ \\
		$\phi_1$ & $0.998 $ & $0.9249 $ & $0.0333$ & $0.0386$ & & $0.9555 $ & $0.0190$ & $0.0208$ \\
		$\xi   $ & $0.4   $ & $0.7284 $ & $1.1962$ & $1.3040$ & & $0.6963 $ & $0.2206$ & $0.3083$ \\
		\bottomrule
	\end{tabular}%
	\caption{\label{tab:KF_T10}Parameter estimates for the discretized version of~\eqref{eq:SDE_CKLS}, as in~\eqref{eq:SDE_CKLS_dis}, for the Kalman Filter algorithm with a mixture of two (left) and seven (right) Gaussian distributions. A thousand simulations were carried out with weekly data and $n=520$, which corresponds to $T=10$ years.}%
\end{table}%

\begin{table}[H]
	\centering
	\begin{tabular}{lrrrrrrrr}
		&       & \multicolumn{3}{c}{\textbf{Kalman Filter (2)}} &       & \multicolumn{3}{c}{\textbf{Kalman Filter (7)}} \\
		\midrule
		& \multicolumn{1}{c}{True} & \multicolumn{1}{c}{Mean} & \multicolumn{1}{c}{Var} & \multicolumn{1}{c}{MSE} &       & \multicolumn{1}{c}{Mean} & \multicolumn{1}{c}{Var} & \multicolumn{1}{c}{MSE} \\
		\midrule
		$\alpha$ & $0.04  $ & $0.0567 $ & $0.0006$ & $0.0008$ & & $0.0567 $ & $0.0006$ & $0.0008$ \\
		$\beta $ & $0.6   $ & $0.8508 $ & $0.1258$ & $0.1887$ & & $0.8508 $ & $0.1258$ & $0.1887$ \\
		$\gamma$ & $1.5   $ & $1.5248 $ & $0.0072$ & $0.0078$ & & $1.4983 $ & $0.0069$ & $0.0069$ \\
		$\phi_0$ & $-0.013$ & $-0.1228$ & $0.0996$ & $0.1116$ & & $-0.0753$ & $0.0299$ & $0.0337$ \\
		$\phi_1$ & $0.998 $ & $0.9825 $ & $0.0018$ & $0.0021$ & & $0.9892 $ & $0.0006$ & $0.0006$ \\
		$\xi   $ & $0.4   $ & $0.5865 $ & $0.0940$ & $0.1288$ & & $0.6672 $ & $0.0507$ & $0.1221$ \\
		\bottomrule
	\end{tabular}%
	\caption{\label{tab:KF_T20}Simulation for weekly data and $n=1040$ ($T=20$ years).}%
\end{table}%

\begin{table}[H]
	\centering
	\begin{tabular}{lrrrrrrrr}
		&       & \multicolumn{3}{c}{\textbf{Kalman Filter (2)}} &       & \multicolumn{3}{c}{\textbf{Kalman Filter (7)}} \\
		\midrule
		& \multicolumn{1}{c}{True} & \multicolumn{1}{c}{Mean} & \multicolumn{1}{c}{Var} & \multicolumn{1}{c}{MSE} &       & \multicolumn{1}{c}{Mean} & \multicolumn{1}{c}{Var} & \multicolumn{1}{c}{MSE} \\
		\midrule
		$\alpha$ & $0.04  $ & $0.0485 $ & $0.0002$ & $0.0003$ & & $0.0485 $ & $0.0002$ & $0.0003$ \\
		$\beta $ & $0.6   $ & $0.7281 $ & $0.0509$ & $0.0673$ & & $0.7281 $ & $0.0509$ & $0.0673$ \\
		$\gamma$ & $1.5   $ & $1.5184 $ & $0.0045$ & $0.0049$ & & $1.4939 $ & $0.0057$ & $0.0058$ \\
		$\phi_0$ & $-0.013$ & $-0.0445$ & $0.0013$ & $0.0022$ & & $-0.0360$ & $0.0008$ & $0.0013$ \\
		$\phi_1$ & $0.998 $ & $0.9937 $ & $2.5\times 10^{-5}$ & $4.4\times 10^{-5}$ & & $0.9949 $ & $1.7\times 10^{-5}$ & $2.7\times 10^{-5}$ \\
		$\xi   $ & $0.4   $ & $0.5169 $ & $0.0208$ & $0.0345$ & & $0.6826 $ & $0.0245$ & $0.1043$ \\
		\bottomrule
	\end{tabular}%
	\caption{\label{tab:KF_T40}Simulation for weekly data and $n=2080$ ($T=40$ years).}%
\end{table}%

\subsubsection{CKLS with stochastic volatility and correlated errors}

We incorporate leverage effect to the CKLS-OU model in~\eqref{eq:SDE_CKLS} by considering correlated Wiener processes, given by
\begin{equation} \label{eq:SDEcor}
\begin{aligned}
\dif r_t &= (\alpha - \beta r_t) \dif t + \sigma_t r_t^\gamma \dif W_{1,t}, \\
\dif \log \sigma_t^2 &= (\theta_0 - \theta_1 \log \sigma_t^2 )\dif t + \xi \dif W_{2,t}, \\
\dif W_{1,t} \dif W_{2,t} &= \rho \dif t.
\end{aligned}
\end{equation}
with $\rho \in [-1,1]$. 

Let $e_{t_i}$ be the residuals from the linear regression, $e_{t_i} = \left[r_{t_i} - \alpha\Delta + (1-\beta\Delta)r_{t_{i-1}}\right]/\sqrt{\Delta} = \sigma_{t_{i_1}}r_{t_{i_1}}^\gamma \varepsilon_{1,t_i}$ with $\varepsilon_{1,t_i}$ standard Gaussian distributed, the discretized version of~\eqref{eq:SDEcor}, setting $y_{t_i} = \log e_{t_i}^2$ and $h_{t_i}=\log \sigma_{t_i}^2$, is
\begin{equation} \label{eq:SDEcor_disc}
\begin{aligned}
y_{t_i} &= h_{t_{i-1}} + 2\gamma \log r_{t_{i-1}} + v_{t_i}, \qquad &&v_{t_i} \sim \log \chi^2, \\
h_{t_i} &= \phi_0 + \phi_1 h_{t_{i-1}} + w_{t_i},            \qquad &&w_{t_i} \sim N(0, \sigma_w^2),
\end{aligned}
\end{equation}
where $\phi_0 = \theta_0 \Delta$, $\phi_1 = 1 - \theta_1 \Delta$, $w_{t_i} = \xi\sqrt{\Delta}\varepsilon_{2,t_i}$, $\varepsilon_{2,t_i}\sim N(0,1)$, and $\sigma_w^2 = \Delta \xi^2$, with $i = 0,1,\dots,n-1$. But because of the logarithmic and square transformation ($v_{t_i} = \log \varepsilon_{1,t_i}^2$), we can not retain the correlation between $\varepsilon_{1,t_i}$ and $w_{t_i}$, that is, $\mathrm{Cor} \lbrace \varepsilon_{1,t_i}, \varepsilon_{2,t_i} \rbrace = \rho$. To overcome this problem, \cite{artigas2004efficient} proposed maintaining the leverage effect by defining
%
$\eta_{t_i} = \rho w_{t_i} + \tilde{\eta}_{t_i}$,
%
where $\tilde{\eta}_{t_i}$ is a normal random variable independent of $\varepsilon_{1,t_i}$ and $\mathbb{V}\textnormal{ar} \left[ \tilde{\eta}_{t_i} \right] = \sigma_w^2(1-\rho)^2$. Note that
\begin{equation*}
\begin{pmatrix} \varepsilon_{1,t_i} \\ \eta_{t_i} \end{pmatrix} \sim N(\mu, \Sigma),
\end{equation*}
where
\begin{equation*}
\mu = \begin{pmatrix}0\\ 0\end{pmatrix}, \qquad
\Sigma = \begin{pmatrix} 1 & \rho\sigma_w \\
\rho\sigma_w & \sigma_w^2 \end{pmatrix},
\end{equation*}
therefore, $\Sigma_{\eta_{t_i} \mid \varepsilon_{1,t_i}} = \sigma_w^2 - \rho^2\sigma_w^2$. The state-space equations for the Kalman Filter algorithm can be written as
\begin{equation} \label{eq:state_space_cor}
\begin{aligned}
y_{t_i} &= h_{t_{i-1}} + 2\gamma \log r_{t_{i-1}} + v_{t_i}, \\
h_{t_i} &= \phi_0 + \phi_1 h_{t_{i-1}} + \eta_{t_i}, 
\end{aligned}
\end{equation}
with $\eta_{t_i} = \rho\sigma_w^2 \varepsilon_{1,t_i} + \tilde{\eta}_{t_i}$ and $\tilde{\eta}_{t_i} \sim N\big(0,  \sigma_w^2(1-\rho)^2 \big)$. Substituting $\varepsilon_{1,t_i}= e_{t_i} \exp\big( -h_{t_{i-1}}/2 \big) r_{t_{i-1}}^{-\gamma} $ in the state equation in~\eqref{eq:state_space_cor}, we have the modified state equation
\begin{equation*}
h_{t_i} = \phi_0 + \phi_1 h_{t_{i-1}} + \rho \,\sigma_w e_{t_i} \exp\bigg( -\frac{h_{t_{i-1}}}{2} \bigg) r_{t_{i-1}}^{-\gamma} + \tilde{\eta}_{t_i} = G(h_{t_{i-1}}) + \tilde{\eta}_{t_i}.
\end{equation*}
This equation is nonlinear, meaning that the filtering equations in~\eqref{eq:KF_alg1}--\eqref{eq:KF_alg2} are no longer applicable. However, \cite{artigas2004efficient} proposed approximating the system by using a time-varying linear Kalman Filter, thus, we modify 
\begin{equation*}
P_{t_{i+1} \mid t_{i}} = g\big(h_{t_{i} \mid t_{i}}\big)^2 P_{t_{i} \mid t_{i}} + \sigma_w^2(1-\rho^2),
\end{equation*}
where
\begin{equation*}
g\big(h_{t_{i} \mid t_{i}}\big) = \frac{\partial G(x)}{\partial x \mathrel{\big|}_{x=h_{t_{i} \mid t_{i}}}} = 
\phi_1 - \frac{1}{2} \rho \, \sigma_w e_{t_{i+1}} \exp\bigg( -\frac{h_{t_i}}{2} \bigg) r_{t_{i}}^{-\gamma}.
\end{equation*}
We used this modification of the filtering equations to do a simulation study for the CKLS-OU model with leverage effect, introduced in~\eqref{eq:SDEcor}. We generated a thousand random sample paths $\lbrace r_{i\Delta}\rbrace_{i=0}^n$, discarding the first 1000 observations as a burn-in period, with weekly frequency for sample sizes $n \in \lbrace 520, 1040, 2080 \rbrace$, that is $T = \lbrace 10, 20, 40 \rbrace$ years, and parameters $\boldsymbol{\theta} = (\alpha, \beta, \gamma, \phi_0, \phi_1, \xi, \rho)' = (0.04, 0.6, 1.5, -0.010, 0.998, 0.4, -0.5)'$. 

Tables~\ref{tab:KF_cor_T10}--\ref{tab:KF_cor_T40} contain the mean, variance and mean squared error for the estimations obtained using the Kalman Filter with a mixture of two (left) and seven (right) Gaussian distributions, for $T=10, 20$ and $40$ years, respectively. Both methods perform similarly, achieving the seven mixture method slightly more accurate estimations, as the MSE is lower. However, they perform poorly when estimating the correlation $\rho$, even with 40 years of weekly data.

\begin{table}[H]
	\centering
	\begin{tabular}{lrrrrrrrr}
		&       & \multicolumn{3}{c}{\textbf{Kalman Filter (2)}} &       & \multicolumn{3}{c}{\textbf{Kalman Filter (7)}} \\
		\midrule
		& \multicolumn{1}{c}{True} & \multicolumn{1}{c}{Mean} & \multicolumn{1}{c}{Var} & \multicolumn{1}{c}{MSE} &       & \multicolumn{1}{c}{Mean} & \multicolumn{1}{c}{Var} & \multicolumn{1}{c}{MSE} \\
		\midrule
		$\alpha$ & 0.04  & 0.0542 & 0.0007 & 0.0009 &       & 0.0542 & 0.0007 & 0.0009 \\
		$\beta $ & 0.6   & 0.8127 & 0.1600 & 0.2053 &       & 0.8127 & 0.1600 & 0.2053 \\
		$\gamma$ & 1.5   & 1.5812 & 0.0560 & 0.0626 &       & 1.5350 & 0.0443 & 0.0455 \\
		$\phi_0$ & -0.010 & -0.3974 & 2.2753 & 2.4257 &       & -0.2071 & 2.0415 & 2.0805 \\
		$\phi_1$ & 0.998 & 0.9330 & 0.1030 & 0.1072 &       & 0.9839 & 0.0584 & 0.0586 \\
		$\xi   $ & 0.4   & 0.6656 & 0.8315 & 0.9021 &       & 0.5251 & 0.2638 & 0.2795 \\
		$\rho  $ & -0.5  & -0.1697 & 0.1295 & 0.2386 &       & -0.1855 & 0.0854 & 0.1843 \\
		\bottomrule
	\end{tabular}%
	\caption{\label{tab:KF_cor_T10} Parameter estimates for the discretized version of~\eqref{eq:SDEcor}, as in~\eqref{eq:SDEcor_disc}, for the Kalman Filter algorithm with a mixture of two (left) and seven (right) Gaussian distributions, with weekly data and $n=520$, which corresponds to $T=10$ years.}%
\end{table}%

\begin{table}[H]
	\centering
	\begin{tabular}{lrrrrrrrr}
		&       & \multicolumn{3}{c}{\textbf{Kalman Filter (2)}} &       & \multicolumn{3}{c}{\textbf{Kalman Filter (7)}} \\
		\midrule
		& \multicolumn{1}{c}{True} & \multicolumn{1}{c}{Mean} & \multicolumn{1}{c}{Var} & \multicolumn{1}{c}{MSE} &       & \multicolumn{1}{c}{Mean} & \multicolumn{1}{c}{Var} & \multicolumn{1}{c}{MSE} \\
		\midrule
		$\alpha$ & 0.04  & 0.0485 & 0.0004 & 0.0004 &       & 0.0485 & 0.0004 & 0.0004 \\
		$\beta $ & 0.6   & 0.7282 & 0.0836 & 0.1000 &       & 0.7282 & 0.0836 & 0.1000 \\
		$\gamma$ & 1.5   & 1.6027 & 0.0827 & 0.0933 &       & 1.5392 & 0.0327 & 0.0343 \\
		$\phi_0$ & -0.010 & -0.2130 & 1.8248 & 1.8662 &       & -0.0335 & 0.5314 & 0.5320 \\
		$\phi_1$ & 0.998 & 0.9842 & 0.0355 & 0.0357 &       & 1.0155 & 0.0171 & 0.0174 \\
		$\xi   $ & 0.4   & 0.5286 & 0.2646 & 0.2811 &       & 0.5591 & 0.1389 & 0.1642 \\
		$\rho  $ & -0.5  & -0.1418 & 0.0694 & 0.4813 &       & -0.1742 & 0.0238 & 0.4784 \\
		\bottomrule
	\end{tabular}%
	\caption{\label{tab:KF_cor_T20}Simulation for weekly data and $n=1040$ ($T=20$ years).}%
\end{table}%

\begin{table}[H]
	\centering
	\begin{tabular}{lrrrrrrrr}
		&       & \multicolumn{3}{c}{\textbf{Kalman Filter (2)}} &       & \multicolumn{3}{c}{\textbf{Kalman Filter (7)}} \\
		\midrule
		& \multicolumn{1}{c}{True} & \multicolumn{1}{c}{Mean} & \multicolumn{1}{c}{Var} & \multicolumn{1}{c}{MSE} &       & \multicolumn{1}{c}{Mean} & \multicolumn{1}{c}{Var} & \multicolumn{1}{c}{MSE} \\
		\midrule
		$\alpha$ & 0.04  & 0.0452 & 0.0002 & 0.0003 &       & 0.0452 & 0.0002 & 0.0003 \\
		$\beta $ & 0.6   & 0.6781 & 0.0502 & 0.0563 &       & 0.6781 & 0.0502 & 0.0563 \\
		$\gamma$ & 1.5   & 1.5675 & 0.1212 & 0.1257 &       & 1.5232 & 0.0336 & 0.0342 \\
		$\phi_0$ & -0.010 & -0.2519 & 2.3199 & 2.3785 &       & -0.0225 & 0.7764 & 0.7766 \\
		$\phi_1$ & 0.998 & 0.9776 & 0.0254 & 0.0258 &       & 1.0154 & 0.0120 & 0.0123 \\
		$\xi   $ & 0.4   & 0.5446 & 0.1866 & 0.2075 &       & 0.5809 & 0.0767 & 0.1094 \\
		$\rho  $ & -0.5  & -0.2564 & 0.0925 & 0.6647 &       & -0.2469 & 0.0588 & 0.6167 \\
		\bottomrule
	\end{tabular}%
	\caption{\label{tab:KF_cor_T40}Simulation for weekly data and $n=2080$ ($T=40$ years).}%
\end{table}%

\section{A GoF test for diffusion processes} \label{sec:3}

In this section, two goodness-of-fit test are introduced. Based on the methodology developed by~\cite{stute1997nonparametric} and extending the goodness-of-fit test presented in~\cite{monsalve2011goodness}, we propose a test for the parametric form of the drift and diffusion functions of the continuous-time stochastic volatility models in~\eqref{eq:SVmodel1}--\eqref{eq:SVmodel2}. The test for the drift function is based on the integrated regression function of the process, while the test for diffusion function relies on the integrated volatility function. The test statistics are based on a distance of the resulting residual marked empirical processes from their expected zero mean, measured by Kolmogorov-Smirnov and Cramér-von Mises functionals. In both tests, the distribution of the statistic is approximated by bootstrap techniques.

\subsection{Test for the volatility function} \label{sec:gof_vola}

The goodness-of-fit test for the parametric form of the volatility function $\nu_1(\cdot)$ in~\eqref{eq:SVmodel1}--\eqref{eq:SVmodel2} under the assumption that the null hypothesis
\begin{equation*}
\mathcal{H}_{0\nu} \colon \nu_1 \in \lbrace \nu_1 (\cdot, \boldsymbol{\theta})\colon \boldsymbol{\theta} \in \Theta \rbrace
\end{equation*}
holds, is based on the integrated conditional variance function
\begin{equation*}
V_0(r,x) = \int_{-\infty}^{r} \int_{-\infty}^{x} \nu_1^2 (u, v) \dif F_{\boldsymbol{\theta}} (u,v) = 
\mathbb{E} \big[ \nu_1^2 (r_t, \boldsymbol{\theta}) \mathbbm{1}_{\lbrace r_{t} \leq r, \sigma^2_{t} \leq x \rbrace} \big],
\end{equation*}
with $r, x \in \mathbb{R}$ and where $F$ is the stationary distribution of $\lbrace r_t, \sigma_t^2 \rbrace$ and $\mathbbm{1}_{\lbrace \cdot \rbrace}$ the indicator function. Considering an empirical estimator of $V(r,x)$,
\begin{equation*}
V_{0n} (r,x) = \frac{1}{n} \sum_{i=0}^{n-1} \mathbbm{1}_{\lbrace r_{t_i} \leq r, \sigma^2_{t_i} \leq x \rbrace} \left( \frac{r_{t_{i+1}} - r_{t_i}}{\Delta} - m_1(r_{t_i}, \hat{\boldsymbol{\theta}})\right)^2,
\end{equation*}
and assuming that $\hat{\boldsymbol{\theta}}$ is an root-$n$ consistent estimator of the true parameter $\boldsymbol{\theta}$, the goodness-of-fit test is based on the empirical process
\begin{equation*}
R_n^\nu (r,x) = \frac{1}{\sqrt{n}} \sum_{i=0}^{n-1} \mathbbm{1}_{\lbrace r_{t_i} \leq r, \hat{\sigma}^2_{t_i} \leq x \rbrace} \left[  \left( \frac{r_{t_{i+1}} - r_{t_i}}{\Delta} - m_1(r_{t_i}, \hat{\boldsymbol{\theta}})\right)^2 - \frac{\hat{\sigma}^2_{t_i} \nu_1^2 (r_{t_i}, \hat{\boldsymbol{\theta}})}{\Delta} \right], 
\end{equation*}
with $r, x \in \mathbb{R}$ and $\hat{\sigma}^2_t$ an estimate of the volatility. A continuous functional $\Psi(\cdot)$ of the empirical process can be considered to define the test statistic $U_n = \Psi(R_n^\nu)$. The null hypothesis is rejected if $U_n > c_{1-\alpha}$, where $c_{1-\alpha}$ is the critical value for the $\alpha$-level test,
\begin{equation*}
\mathbb{P} \left( \Psi(R_n^\nu) > c_{1-\alpha} \mid \mathcal{H}_{0\nu} \right) = \alpha.
\end{equation*}
The critical value $c_{1-\alpha}$ can be determined by approximating the distribution of the process $R_n^\nu$ using bootstrap techniques \citep{stute1998bootstrap}. Let $c^*_{1-\alpha}$ denote the bootstrap approximated critical value $c_{1-\alpha}$, so that $\mathbb{P}^* \left( \Psi(R_n^{\nu*}) > c^*_{1-\alpha} \mid \mathcal{H}_{0\nu} \right) = \alpha$, where $\mathbb{P}^*$ is the probability measure generated by the bootstrap sample and the bootstrap counterpart of the empirical process $R_n^\nu (\cdot)$ is given by
\begin{equation*}
R_n^{\nu*} (r,x) = \frac{1}{\sqrt{n}} \sum_{i=0}^{n-1} \mathbbm{1}_{\lbrace r_{t_i}^* \leq r, \hat{\sigma}^{*2}_{t_i} \leq x \rbrace} \left[  \left( \frac{r_{t_{i+1}}^* - r_{t_i}^*}{\Delta} - m_1(r_{t_i}^*, \hat{\boldsymbol{\theta}}^*)\right)^2 - \frac{\hat{\sigma}^{*2}_{t_i} \nu_1^2 (r_{t_i}^*, \hat{\boldsymbol{\theta}}^*)}{\Delta} \right], 
\end{equation*}
with $\hat{\boldsymbol{\theta}}^*$ an estimator obtained from the bootstrapped sample $\lbrace (r_{t_i}^*, \sigma_{t_i}^{*2}) \rbrace$ (the procedure to obtain the resamples is defined in Section~\ref{sec:boot}). 
Bootstrap replicates of the statistic $U_n^{*j} = \Psi \big(R_n^{\nu*} (r,x)\big)$, for $j=1,\dots,B$, are obtained and, using Monte Carlo techniques, the critical value is approximated with the statistic of order $\lceil B(1-\alpha) \rceil$ from the $B$ bootstrap replicates, that is, $c^*_{1-\alpha} = U_n^{*\lceil B(1-\alpha) \rceil}$ (see Algorithm~\ref{alg:boot} for a summary of the bootstrap procedure to approximate the critical value $c_{1-\alpha}$). The null hypothesis $\mathcal{H}_{0\nu}$ is rejected if $U_n > c^*_{1-\alpha}$. 

The functional $\Psi(\cdot)$ can take the form of the Kolmogorov-Smirnov (KS) and Cramér-von Mises (CvM) criteria, so that the statistics can be expressed as
\begin{equation*}
\begin{aligned}
U_n^{KS} &\coloneqq \sup\limits_{r,x} \left| R_n^\nu (r,x)\right|, \\
U_n^{CvM} &\coloneqq \int \int_{\mathbb{R}^2} \left( R_n^\nu (r,x) \right)^2 F_n (\dif r, \dif x), 
\end{aligned}
\end{equation*}
respectively, where $F_n$ is the empirical distribution of $\lbrace r_{t_i}, \hat{\sigma}^2_{t_i} \rbrace_{i=0}^{n-1}$. The empirical $p$-value is estimated with the proportion of the B bootstrap replicates $U_n^{*j}$ exceeding $U_n$, that is,
\begin{equation*}
\frac{\sharp \lbrace U_n^{*j} > U_n \rbrace}{B}.
\end{equation*}

\subsection{Test for the drift function} \label{sec:gof_drift}

Aiming to test if the parametric form of the drift function $m_1 (\cdot)$ in~\eqref{eq:SVmodel1}--\eqref{eq:SVmodel2} belongs to a certain parametric family, we establish the null hypothesis
\begin{equation*}
\mathcal{H}_{0m} \colon m_1 \in \lbrace m_1(\cdot, \boldsymbol{\theta})\colon \boldsymbol{\theta} \in \Theta \rbrace.
\end{equation*}
We propose a test based on the integrated regression function $I(x) = \mathbb{E} \left[ Y \mathbbm{1}_{\lbrace X \leq x \rbrace} \right] = \int_{-\infty}^{x} m(y) \dif F(y)$, where $F$ is the marginal distribution function of $X$. An empirical estimator of the integrated regression function for the model~\eqref{eq:SVmodel1}--\eqref{eq:SVmodel2} is given by
\begin{equation*}
I_n (r) = \frac{1}{n} \sum_{i=0}^{n-1} \mathbbm{1}_{\lbrace r_{t_i} \leq r \rbrace} \frac{r_{t_{i+1}} - r_{t_i}}{\Delta},
\end{equation*}
and an estimator under the null hypothesis $\mathcal{H}_{0m}$ is
\begin{equation*}
I_{0n} (r) = \frac{1}{n} \sum_{i=0}^{n-1} \mathbbm{1}_{\lbrace r_{t_i} \leq r \rbrace}  m_1(r_{t_i}, \hat{\boldsymbol{\theta}}).
\end{equation*}
The goodness-of-fit test compares the estimated integrated regression function $I_n (\cdot)$ with the estimation obtained under the null hypothesis, that is, $I_{0n} (\cdot)$. Therefore, the test is defined by the process
\begin{equation*}
R_n^m (r) = \frac{1}{\sqrt{n}} \sum_{i=0}^{n-1} \mathbbm{1}_{\lbrace r_{t_i} \leq r \rbrace} \left( \frac{r_{t_{i+1}} - r_{t_i}}{\Delta} - m_1(r_{t_i}, \hat{\boldsymbol{\theta}}) \right), 
\end{equation*}
with $r \in \mathbb{R}$ and where $\hat{\boldsymbol{\theta}}$ is a $\sqrt{n}$-consistent estimator of the true parameter vector $\boldsymbol{\theta}$. As in the previous test, a continuous functional $\Psi (\cdot)$ can be considered to define the statistic $T_n = \Psi (R_n^m)$, such as the Kolmogorov-Smirnov and Cramér-von Mises criteria,
\begin{equation*}
T_n^{KS} \coloneqq \sup\limits_{r} \left| R_n^m (r)\right| \qquad \text{and} \qquad T_n^{CvM} \coloneqq \int_{\mathbb{R}}  R_n^m (r)^2 F_n (\dif r), 
\end{equation*}
respectively, with $F_n$ the empirical distribution function of $\lbrace r_{t_i} \rbrace_{i=0}^{n-1}$. The null hypothesis is rejected if $T_n > c_{1-\alpha}$. Again, the critical value $c_{1-\alpha}$ can be approximated by its bootstrap counterpart $c^*_{1-\alpha}$, such that $\mathbb{P}^* \left( \Psi(R_n^{m*}) > c^*_{1-\alpha} \mid \mathcal{H}_{0m} \right) = \alpha$ with
\begin{equation*}
R_n^{m*} (r) = \frac{1}{\sqrt{n}} \sum_{i=0}^{n-1} \mathbbm{1}_{\lbrace r_{t_i} \leq r \rbrace} \left( u^*_{t_i} - m_1(r_{t_i}, \hat{\boldsymbol{\theta}}^*) \right), 
\end{equation*}
where $u_{t_i} = (r_{t_{i+1}} - r_{t_i}) / \Delta$ and $u_{t_i}^* = m_1 (r_{t_i}, \hat{\boldsymbol{\theta}}) + e_{t_i}^*$ is obtained sampling the innovations $e_{t_i}$ with the algorithm introduced in Section~\ref{sec:boot}, and the bootstrap estimator $\hat{\boldsymbol{\theta}}^*$ is calculated from the bootstrap sample $\lbrace (r_{t_i}^*, \sigma_{t_i}^{*2}) \rbrace$. The approximated critical value $c^*_{1-\alpha}$ is achieved by means of Monte Carlo techniques, that is, $c^*_{1-\alpha} = T_n^{*\lceil B(1-\alpha) \rceil}$, with the order $\lceil B(1-\alpha) \rceil$ statistic from $B$ bootstrap replicates $\big\lbrace T_n^{*j} = \Psi(R_n^{m*}) \big\rbrace_{j=1}^B$. As explained in the previous test, the empirical $p$-value is the proportion of the $B$ bootstrap replicates $T_n^{*j}$ exceeding $T_n$, $\sharp \lbrace T_n^{*j} > T_n \rbrace / B$.

\subsection{Bootstrap resampling procedure} \label{sec:boot}

The estimation of $\boldsymbol{\theta}$ and the bootstrap sample $\lbrace (r_{t_i}^*, \sigma_{t_i}^{*2}) \rbrace$ can be obtained with the Kalman Filter algorithm \citep{shumway2000time}. Given the filtering equations introduced in~\eqref{eq:KF_alg1}--\eqref{eq:KF_alg2}, the bootstrap resample algorithm for model~\eqref{eq:SVmodel1}--\eqref{eq:SVmodel2} can be implemented as follows. First, calculate the residuals from the linear regression $e_{t_i} = [r_{t_i} - m_1(r_{t_i}, \boldsymbol{\theta})] / \sqrt{\Delta} = \sigma_{t_i} \nu_1(r_{t_i}, \boldsymbol{\theta}) \varepsilon_{1,t_i}$; second, define $y_{t_i} = \log e_{t_i}^2 \quad \text{and} \quad h_{t_i} = \log \sigma_{t_i}^2$ to linearize the space equation, therefore we have the following space-state equations,
\begin{alignat*}{2}
y_{t_i} &= h_{t_i} + \log \nu_1(r_{t_i}, \boldsymbol{\theta})^2 + v_{t_i}, \qquad\quad &&v_{t_i} \sim \log \chi^2_1, \\
h_{t_i} &= m_2(h_{t_{i}}, \boldsymbol{\theta}) + w_{t_i}, \qquad\quad &&w_{t_i} \sim N(0, \sigma_w^2),
\end{alignat*}
with $w_{t_i} = \nu_2 (h_{t_i}, \boldsymbol{\theta}) \sqrt{\Delta}\,\varepsilon_{2,t_i}$, $\varepsilon_{2,t_i} \sim N(0,1)$ and $\sigma_w^2 = \Delta \nu_2 (h_{t_i}, \boldsymbol{\theta})^2$. Given the filtering equations~\eqref{eq:KF_alg1}--\eqref{eq:KF_alg2} we have
\begin{equation*}
y_{t_i} = h_{t_i\mid {t_{i-1}}} + \log \nu_1 (r_{t_{i-1}}, \boldsymbol{\theta})^2 + \pi_{0,t_i} \epsilon_{0,t_i} + \pi_{1,t_i} (\epsilon_{1,t_i} + \mu_1).
\end{equation*}
Let $\hat{\boldsymbol{\theta}}$ denote the maximum likelihood estimator of the true parameter $\boldsymbol{\theta}$, $\hat{\boldsymbol{\theta}} = \argmax\limits_\Theta \ln \mathcal{L}_Y (\boldsymbol{\theta})$, by means of the Kalman Filter algorithm, and $\epsilon_{j,t_i}$, with $\mathbb{V}\mathrm{ar} \left[ \epsilon_{j,t_i} \right] = \hat{\Sigma}_{j,t_i}$, the innovations and the innovations variance obtained by running the filter under $\hat{\boldsymbol{\theta}}$. To implement the bootstrap algorithm we consider the standardized innovations
\begin{equation*}
\tilde{\epsilon}_{j,t_i} = \hat{\Sigma}_{j,t_i}^{-1/2} \epsilon_{j,t_i}, \quad \text{ for } j = 0,1,
\end{equation*}
and sample with replacement from $\lbrace \tilde{\epsilon}_{j,t_i} \rbrace_{i=0}^{n-1}$ to obtain the bootstrap sample of standardized innovations $\lbrace \tilde{\epsilon}_{j,t_i}^* \rbrace_{i=0}^{n-1}$. We then generate the bootstrap sample $\lbrace (y_{t_i}^*, h_{t_i}^{*}) \rbrace_{i=0}^{n-1}$ with
\begin{equation} \label{eq:boot}
\begin{aligned}
h_{t_{i+1}}^* &= m_2(h_{t_i\mid t_{i-1}}, \hat{\boldsymbol{\theta}}) + \hat{\pi}_{0,t_i} K_{0,t_i} \hat{\Sigma}_{0,t_i}^{1/2} \tilde{\epsilon}_{0,t_i}^* + \hat{\pi}_{1,t_i} K_{1,t_i} \hat{\Sigma}_{1,t_i}^{1/2} \tilde{\epsilon}_{1,t_i}^*, \\
y_{t_i}^*     &= h_{t_i\mid t_{i-1}} + \log \nu_1(r_{t_i}, \hat{\boldsymbol{\theta}})^2 + \hat{\pi}_{1,t_i} \hat{\mu}_1 + \hat{\pi}_{0,t_i} \hat{\Sigma}_{0,t_i}^{1/2} \tilde{\epsilon}_{0,t_i}^* + \hat{\pi}_{1,t_i} \hat{\Sigma}_{1,t_i}^{1/2} \tilde{\epsilon}_{1,t_i}^*,
\end{aligned}
\end{equation}
where $r_{t_i}$ remains fixed and the bootstrapped dataset is given by $\exp (y_{t_i}^*) = e_t^{2*}$. Algorithm~\ref{alg:boot} summarizes the bootstrap procedure to approximate the critical value $c_{1-\alpha}$.
\vspace{0.5\baselineskip}

\begin{algorithm}[Bootstrap resampling procedure] \label{alg:boot} The critical value $c_{1-\alpha}$ can be approximated with the following bootstrap procedure:
	
	\begin{enumerate}
		\item Obtain the maximum likelihood estimator of the true parameter $\boldsymbol{\theta}$, that is, $\hat{\boldsymbol{\theta}} = \argmax\limits_\Theta \ln \mathcal{L}_Y (\boldsymbol{\theta})$, by means of the Kalman Filter algorithm introduced in~\eqref{eq:kf_loglike}.
		\item Construct the bootstrap sample $\lbrace (y_{t_i}^*, \sigma_{t_i}^{*2}) \rbrace_{i=0}^{n-1}$ as in~\eqref{eq:boot}.
		\item Estimate the parameter vector $\hat{\boldsymbol{\theta}}^*$ from the bootstrap resample $\lbrace (y_{t_i}^*, \sigma_{t_i}^{*2}) \rbrace_{i=0}^{n-1}$ obtained in Step~2.
		\item Compute the bootstrap version of the process $R_n^{\nu*} (r,x)$ or $R_n^{m*} (r)$, for $r, x \in \mathbb{R}$.
		\item Determine $U_n^* = \Psi \big(R_n^{\nu*} (r,x) \big)$ or $T_n^* = \Psi \big(R_n^{m*} (r) \big)$.
		\item Repeat $B$ times the previous Steps~2--5 to obtain $j=1,\dots,B$, bootstrap replicates $U_n^{*j}$ or $T_n^{*j}$.
		\item Approximate the critical value $\hat{c}^*_{1-\alpha} = U_n^{*\lceil B(1-\alpha)\rceil}$ or $\hat{c}^*_{1-\alpha} = T_n^{*\lceil B(1-\alpha)\rceil}$.
	\end{enumerate}
\end{algorithm}

\section{Simulation study} \label{sec:4}

In this Section, a simulation study to illustrate the finite sample properties of the tests was conducted under different settings, for both the drift and volatility goodness-of-fit tests. As to our knowledge, the tests available in the literature for continuous-time stochastic volatility models do not test the same null hypothesis as our test, that is, the parametric form of the diffusion function, we are not including a comparison with other procedures.

\subsection{Drift test}

The performance of the goodness-of-fit test for the parametric form of the drift function introduced in Section~\ref{sec:gof_drift}, regarding size and power, is illustrated by means of a simulation study. We test the null hypothesis that the drift function $m_1(\cdot)$ of the SDE in~\eqref{eq:SVmodel1}--\eqref{eq:SVmodel2} belongs to a certain parametric family,
\begin{equation*}
\mathcal{H}_{0m} \colon m_1 \in \lbrace m_1(\cdot,\boldsymbol{\theta}) \colon \boldsymbol{\theta} \in \Theta \rbrace.
\end{equation*}
We consider the CKLS-OU model as the null hypothesis and to evaluate the power of the test we use a series of alternative models indexed by the parameter $\rho$ of the non-linear function $\rho_m (r_t) = \rho (1-r_t^\rho)$,
\begin{align*}
\dif r_t &= \big(\alpha - \beta r_t + \rho (1 - r_t^{\,\rho}) \big)\dif t + \sigma_t r_t^\gamma \dif W_{1,t}, \\ 
\dif \ln \sigma_t^2 &= (\theta_0 - \theta_1 \log \sigma_t^2)\dif t + \xi \dif W_{2,t}, 
\end{align*}
with $\rho \in \lbrace 0, 0.07, 0.09, 0.10, 0.125, 0.15 \rbrace$, where $\rho = 0$ under the null hypothesis. The parameter vector for the data generating process (DGP) is $\boldsymbol{\theta} = (\alpha, \beta, \gamma, \theta_0, \theta_1, \xi)' = (0.04,0.6,1.5,-0.7,0.1,0.4)$. The process was generated with weekly frequency ($\Delta = 1/52$) for different sample sizes $n \in \lbrace 500, 1000, 1500, 2000 \rbrace$ and the first thousand observations were discarded as a burn-in period. The rate of rejection ($\hat{\alpha}$) is calculated based on $1000$ Monte Carlo replicates and $B=1000$ bootstrap resamples (see Algorithm~\ref{alg:boot}) for the Kolmogorov-Smirnov ($\hat{\alpha}_{\text{KS}}$) and Cramér-von Mises $(\hat{\alpha}_{\text{CvM}}$) criteria.

Table~\ref{tab:size_drift} show the size (first row) and power of the goodness-of-fit test for the drift function $m_1(\cdot)$ for the null hypothesis that the drift function follows the CKLS-OU parametric form, that is, $\mathcal{H}_{0m} \colon m_1 (r_t,\boldsymbol{\theta}) = \alpha - \beta r_t$, with $\alpha = 0.05$. Regarding the size, the tests are well calibrated as both rejection rates are very close to the nominal level $\alpha$. The behavior of the power, on the other hand, shows an increase with the sample size, as expected, and the higher the value of $\rho$, the further we depart from the null hypothesis, obtaining higher rejection rates.

\begin{table}[H]
	\centering
	\begin{tabular}{llrrrrcrrrr}
		\toprule
		&& \multicolumn{4}{c}{$\hat{\alpha}_{\text{KS}}$} & & \multicolumn{4}{c}{$\hat{\alpha}_{\text{CvM}}$} \\
		\midrule
		\multicolumn{1}{c}{DGP} && 500 & 1000  & 1500  & 2000 & & 500 & 1000 & 1500 & 2000 \\
		\midrule
		$\rho = $ 0     && 0.047 & 0.043 & 0.045 & 0.044 && 0.054 & 0.051 & 0.060 & 0.056 \\
		$\rho = $ 0.07  && 0.102 & 0.186 & 0.203 & 0.301 && 0.169 & 0.267 & 0.298 & 0.356 \\
		$\rho = $ 0.09  && 0.301 & 0.305 & 0.356 & 0.456 && 0.314 & 0.364 & 0.448 & 0.508 \\
		$\rho = $ 0.10  && 0.365 & 0.456 & 0.481 & 0.526 && 0.441 & 0.528 & 0.560 & 0.606 \\
		$\rho = $ 0.125 && 0.523 & 0.618 & 0.669 & 0.703 && 0.531 & 0.636 & 0.682 & 0.747 \\
		$\rho = $ 0.15  && 0.790 & 0.848 & 0.869 & 0.901 && 0.785 & 0.864 & 0.907 & 0.923 \\
		\bottomrule
	\end{tabular}%
	\caption{\label{tab:size_drift}Size ($\rho = 0$) and power simulation for the CKLS-OU model drift test, with $\alpha=0.05$, for the null hypothesis $\mathcal{H}_0 \colon m_1 (r_t,\boldsymbol{\theta}) = \alpha - \beta r_t$, under different alternative scenarios $m_1 (r_t, \boldsymbol{\theta}) = \big(\alpha - \beta r_t + \rho (1 - r_t^{\,\rho}) \big)$, for $\rho \in \lbrace 0, 0.07, 0.09, 0.10, 0.125, 0.15 \rbrace$.}%
\end{table}%

\subsection{Volatility test}

The study of the finite sample properties of the goodness-of-fit test for the parametric form of the volatility function introduced in Section~\ref{sec:gof_vola} is accomplish with a simulations study, testing both simple and composite null hypotheses. We test the null hypothesis that the diffusion function $\nu_1 (\cdot)$ of the continuous-time model in~\eqref{eq:SVmodel1}--\eqref{eq:SVmodel2} belongs to a certain parametric family, that is,
\begin{equation*}
\mathcal{H}_{0\nu} \colon \nu_1 \in \lbrace \nu_1 (\cdot, \boldsymbol{\theta}) \colon \boldsymbol{\theta} \in \Theta \rbrace,
\end{equation*}
for the composite hypothesis. We consider three different models under the null hypothesis, which are described in Table~\ref{tab:models}, and, to asses the performance of the power of the test, we take the CKLS-OU model and create a series of alternative scenarios by adding a non-linear function to the diffusion function $\nu_1(\cdot)$
\begin{align*}
\dif r_t &= (\alpha - \beta r_t)\dif t + \big[ \sigma_t r_t^\gamma + \rho (1 - r_t^{\,\rho}) \big] \dif W_{1,t}, \\ 
\dif \ln \sigma_t^2 &= (\theta_0 - \theta_1 \log \sigma_t^2)\dif t + \xi \dif W_{2,t}, 
\end{align*}
with $\rho \in \lbrace 0.007, 0.01, 0.02, 0.04 \rbrace$, where $\rho = 0$ under the null hypothesis (scenario \textit{CKLS-OU} in Table~\ref{tab:models}). The processes were generated with weekly frequency ($\Delta = 1/52$) for sample sizes $n \in \lbrace 500, 1000, 1500, 2000 \rbrace$ and the first thousand observations were discarded. The rate of rejection ($\hat{\alpha}$) is calculated based on $1000$ Monte Carlo replicates and $B=1000$ bootstrap resamples (see Algorithm~\ref{alg:boot}) for the Kolmogorov-Smirnov ($\hat{\alpha}_{\text{KS}}$) and Cramér-von Mises $(\hat{\alpha}_{\text{CvM}}$) criteria.

The size and power of the test for the simple null hypothesis will be evaluated using the CKLS-OU model ($\nu_1 (r_t,\boldsymbol{\theta}) = r_t^\gamma$), testing the null hypothesis $\mathcal{H}_{0\nu}\colon \nu_1 (r_t,\boldsymbol{\theta}) = r_t^{1.5}$ for a set of values $\gamma \in \lbrace 1.5, 1.25, 1.0 \rbrace$.

\begin{table}[H]
	\centering
	\begin{tabular}{ccc}
		\toprule 
		\textbf{Scenario} & \textbf{Model} & \textbf{Parameters} \\ 
		\midrule
			\multirow{2}{*}{OU-OU} & 
			$\begin{array}[t]{ r @{{}={}} l }
			\dif r_t & (\alpha - \beta r_t)\dif t + \sigma_t  \dif W_{1,t} \\ 
			\dif \ln \sigma_t^2 & (\theta_0 - \theta_1 \log \sigma_t^2)\dif t + \xi \dif W_{2,t} 
			\end{array}$  & 
			$\begin{array}[t]{ r @{{}={}} l }
			\boldsymbol{\theta} & (\alpha, \beta, \gamma, \theta_0, \theta_1, \xi)' \\ 
			& (0.04,0.6,-0.7,0.1,0.4)'\phantom{1.5,}
			\end{array}$ \\ 
		\midrule
			\multirow{2}{*}{CKLS-null} & 
			$\begin{array}[t]{ r @{{}={}} l }
			\dif r_t & (\alpha - \beta r_t)\dif t + \sigma_t r_t^\gamma \dif W_{1,t} \\ 
			\dif \ln \sigma_t^2 & \xi \dif W_{2,t} \hphantom{ \,+\, (\theta_0 - \theta_1 \log \sigma_t^2)\dif t}
			\end{array}$  & 
			$\begin{array}[t]{ r @{{}={}} l }
			\boldsymbol{\theta} & (\alpha, \beta, \gamma, \xi)'  \\ 
			& (0.04,0.6,1.5,0.4)'\phantom{-0.7,0.1,}
			\end{array}$ \\ 
		\midrule
			\multirow{2}{*}{CKLS-OU}  & 
			$\begin{array}[t]{ r @{{}={}} l }
			\dif r_t & (\alpha - \beta r_t)\dif t + \sigma_t r_t^\gamma \dif W_{1,t} \\ 
			\dif \ln \sigma_t^2 & (\theta_0 - \theta_1 \log \sigma_t^2)\dif t + \xi \dif W_{2,t} 
			\end{array}$  & 
			$\begin{array}[t]{ r @{{}={}} l }
			\boldsymbol{\theta} & (\alpha, \beta, \gamma, \theta_0, \theta_1, \xi)' \\ 
			& (0.04,0.6,1.5,-0.7,0.1,0.4)'
			\end{array}$ \\ 
		\bottomrule 
	\end{tabular} 
	\caption{\label{tab:models} Scenarios under the composite null hypothesis.}
\end{table}

Table~\ref{tab:size} shows the empirical size and power for simple and composite hypotheses, the later under the null hypotheses for the scenarios in Table~\ref{tab:models}. Regarding the simple hypothesis test, the size (first row) is close to the nominal level $\alpha = 0.05$, with a slight over rejection for the smallest sample size scenario, and the power increases with sample size and shows that the test is capable of discriminating between models with different values of $\gamma$. Focusing on the composite null hypothesis, the estimated sizes (first three rows) remain close the the true nominal level, although the smallest sample sizes show some small deviations, and the power increases both with the sample size and the value of $\rho$, which controls the level of noise added to the diffusion function.

\begin{table}[H]
	\centering
	\begin{tabular}{llrrrrcrrrr}
		\toprule
		&& \multicolumn{4}{c}{$\hat{\alpha}_{\text{KS}}$} & & \multicolumn{4}{c}{$\hat{\alpha}_{\text{CvM}}$} \\
		\midrule
		\multicolumn{1}{c}{$\mathcal{H}_0$} & \multicolumn{1}{c}{DGP} & 500 & 1000  & 1500  & 2000 & & 500 & 1000 & 1500 & 2000 \\
		\midrule
		\multicolumn{11}{c}{Simple hypothesis}  \\
		$\gamma = 1.50$ & CKLS-OU         & \textbf{0.089} & 0.046 & 0.041 & 0.049 && 0.040 & 0.044 & 0.062 & 0.052 \\ 
		$\gamma = 1.50$ & $\gamma = 1.25$ & 0.454 & 0.613 & 0.708 & 0.808 && 0.462 & 0.602 & 0.767 & 0.818 \\
		$\gamma = 1.50$ & $\gamma = 1.00$ & 0.734 & 0.884 & 0.957 & 0.986 && 0.649 & 0.873 & 0.974 & 0.988 \\
		\multicolumn{11}{c}{Composite hypothesis}  \\
		$\nu_1 (r_t,\boldsymbol{\theta}) = 1$          & OU-OU     & 0.037 & 0.046 & 0.046 & 0.058 & & 0.045 & 0.051 & 0.032 & 0.064 \\
		$\nu_1 (r_t,\boldsymbol{\theta}) = r_t^\gamma$ & CKLS-null & \textbf{0.033} & 0.048 & 0.038 & 0.057 & & \textbf{0.035} & 0.051 & 0.034 & 0.063 \\
		$\nu_1 (r_t,\boldsymbol{\theta}) = r_t^\gamma$ & CKLS-OU   & 0.040 & 0.047 & 0.041 & 0.047 & & 0.044 & 0.053 & 0.034 & 0.041 \\
		
		$\nu_1 (r_t,\boldsymbol{\theta}) = r_t^\gamma$ & $\rho = $ 0.007 & 0.125 & 0.177 & 0.206 & 0.331 && 0.099 & 0.148 & 0.190 & 0.303 \\
		$\nu_1 (r_t,\boldsymbol{\theta}) = r_t^\gamma$ & $\rho = $ 0.01  & 0.364 & 0.468 & 0.641 & 0.657 && 0.304 & 0.423 & 0.578 & 0.619 \\
		$\nu_1 (r_t,\boldsymbol{\theta}) = r_t^\gamma$ & $\rho = $ 0.02  & 0.490 & 0.644 & 0.710 & 0.796 && 0.415 & 0.581 & 0.674 & 0.701 \\
		$\nu_1 (r_t,\boldsymbol{\theta}) = r_t^\gamma$ & $\rho = $ 0.04  & 0.527 & 0.766 & 0.870 & 0.951 && 0.491 & 0.620 & 0.714 & 0.819 \\
		\bottomrule
	\end{tabular}%
	\caption{\label{tab:size}Size and power simulation for the volatility function tests, with $\alpha=0.05$, considering simple and composite null hypotheses. Under $\mathcal{H}_0$, rejection rates are boldfaced if they lie outside a 95\%-confidence interval for the nominal level $\alpha$.}%
\end{table}%

\section{Real data applications} \label{sec:5}

\begin{figure}[H]
	\centering
	\begin{subfigure}{.24\textwidth}
		\includegraphics[width=1.05\linewidth]{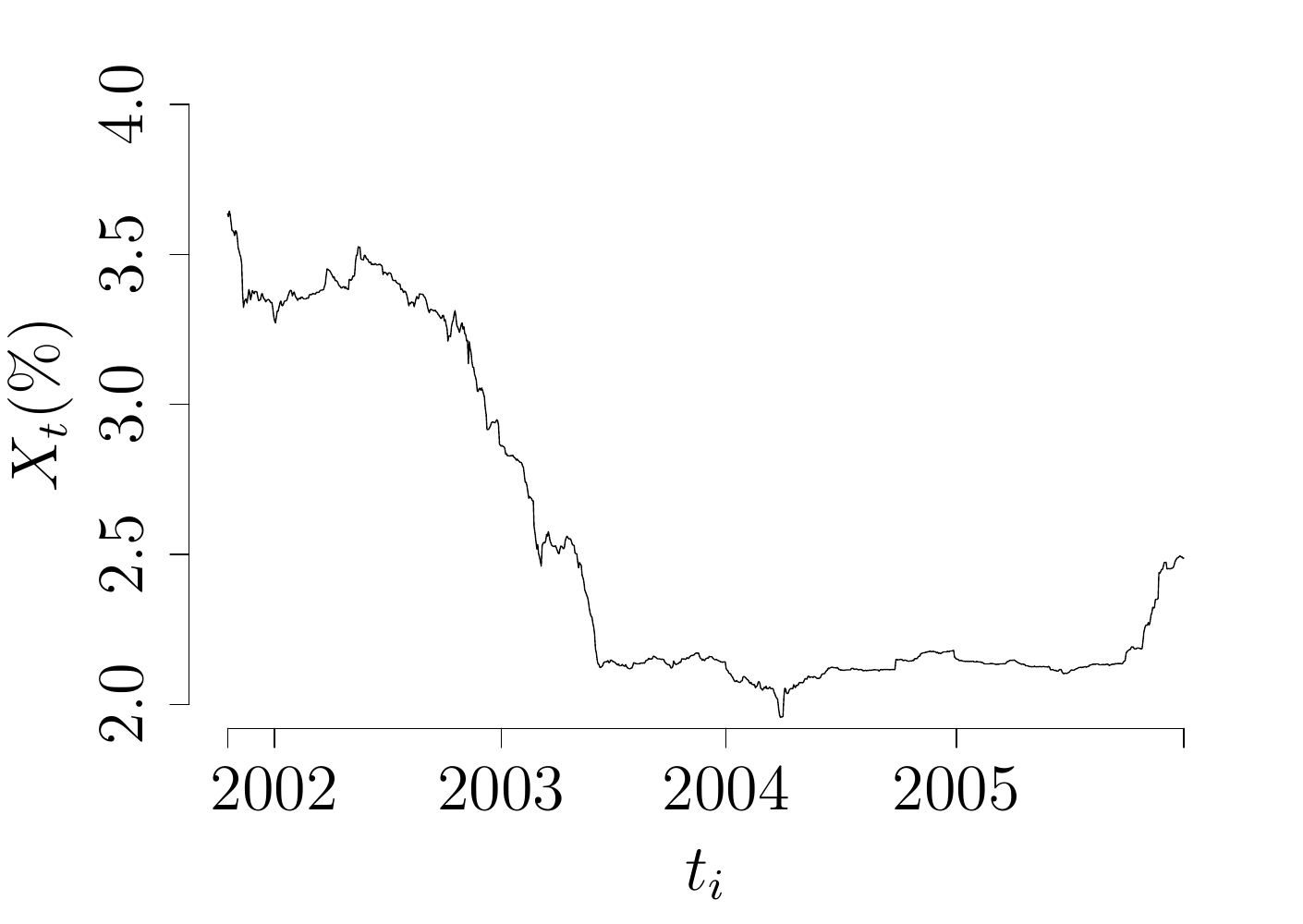}
		\captionsetup{font=footnotesize,textfont=it}
		\caption{Euribor 3 months.}
		\label{fig:euribor3m}
	\end{subfigure}
	\begin{subfigure}{.24\textwidth}
		\includegraphics[width=1.05\linewidth]{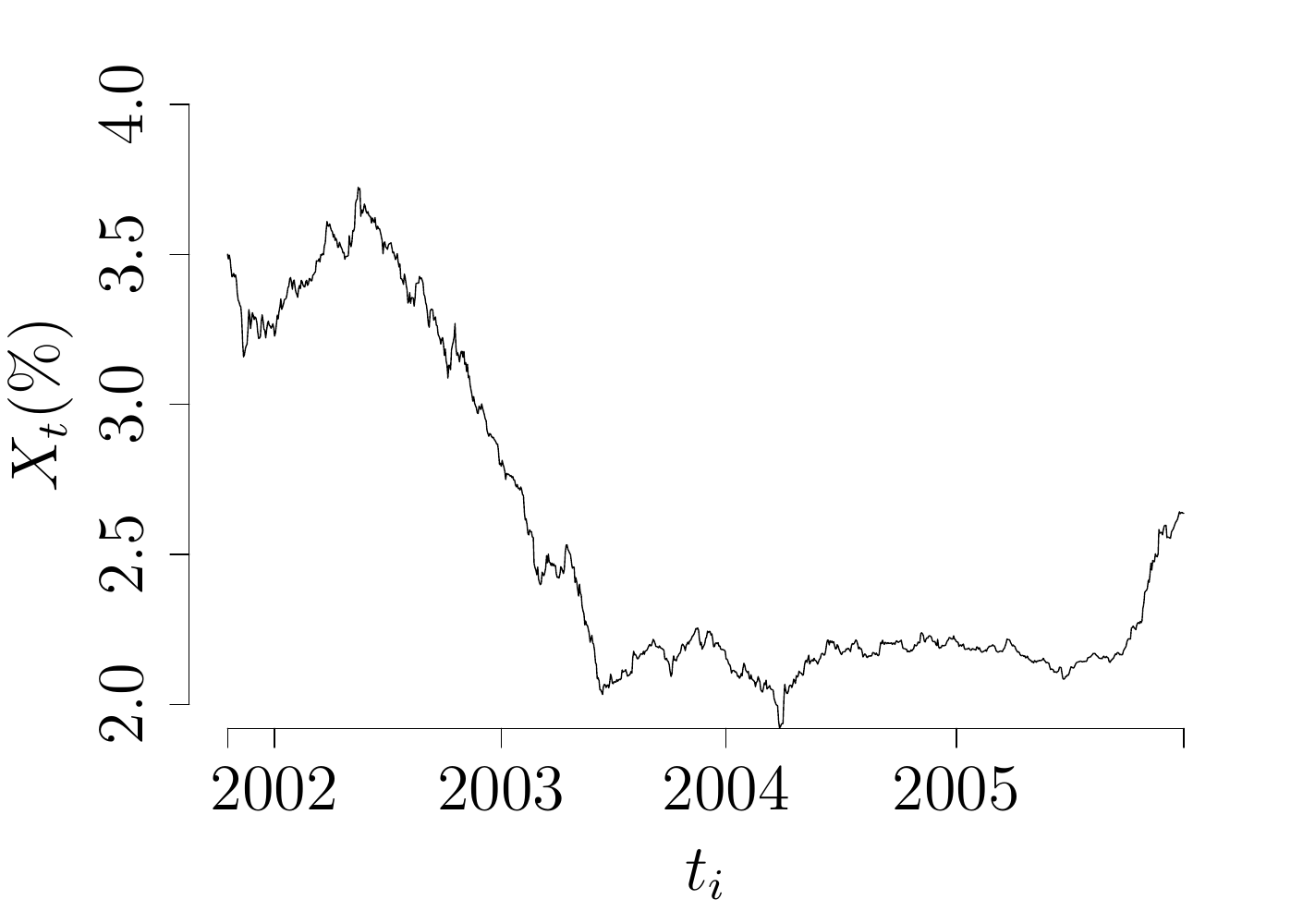}
		\captionsetup{font=footnotesize,textfont=it}
		\caption{Euribor 6 months.}
		\label{fig:euribor6m}
	\end{subfigure}
	\begin{subfigure}{.24\textwidth}
		\includegraphics[width=1.05\linewidth]{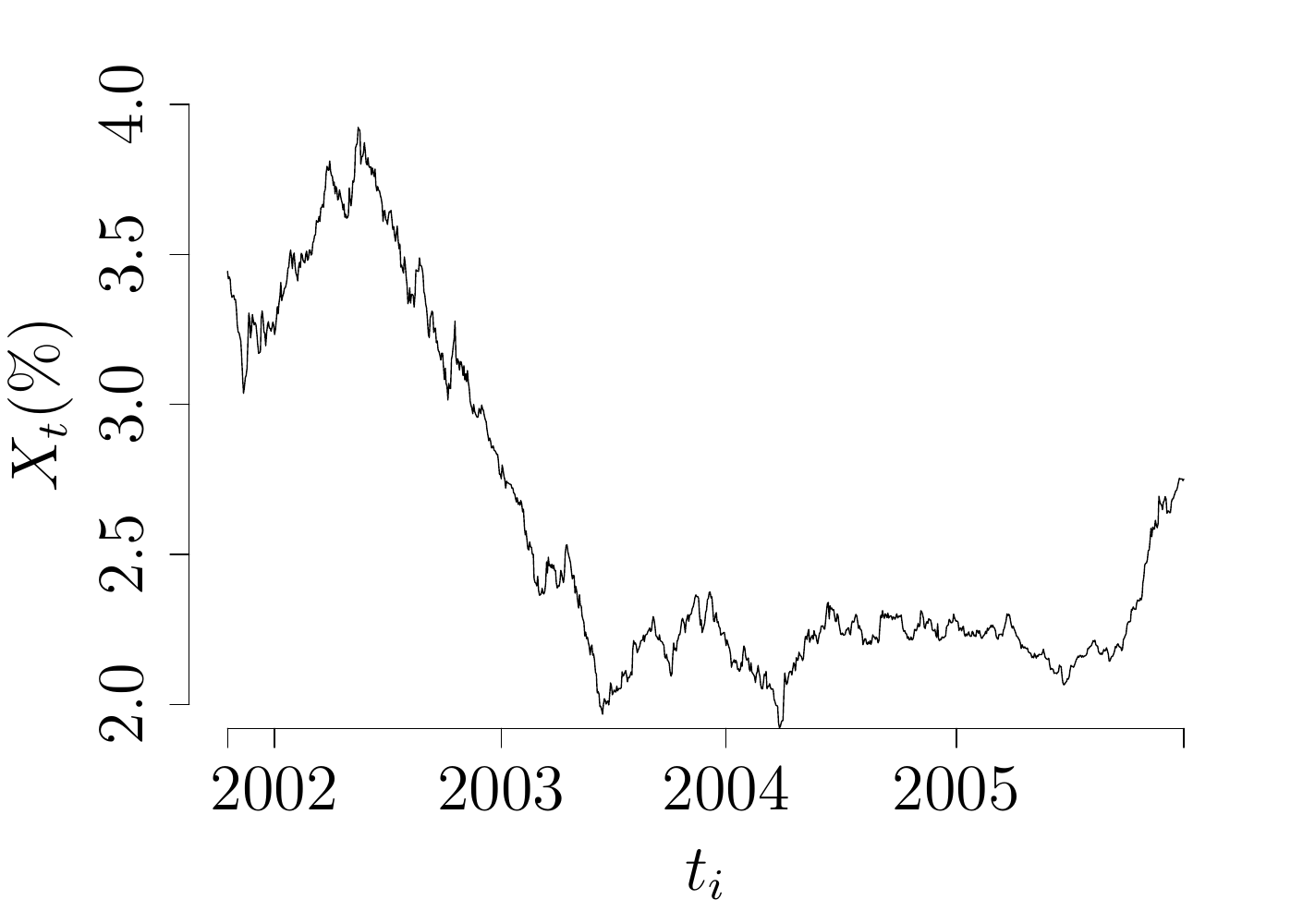}
		\captionsetup{font=footnotesize,textfont=it}
		\caption{Euribor 9 months.}
		\label{fig:euribor9m}
	\end{subfigure}
	\begin{subfigure}{.24\textwidth}
		\includegraphics[width=1.05\linewidth]{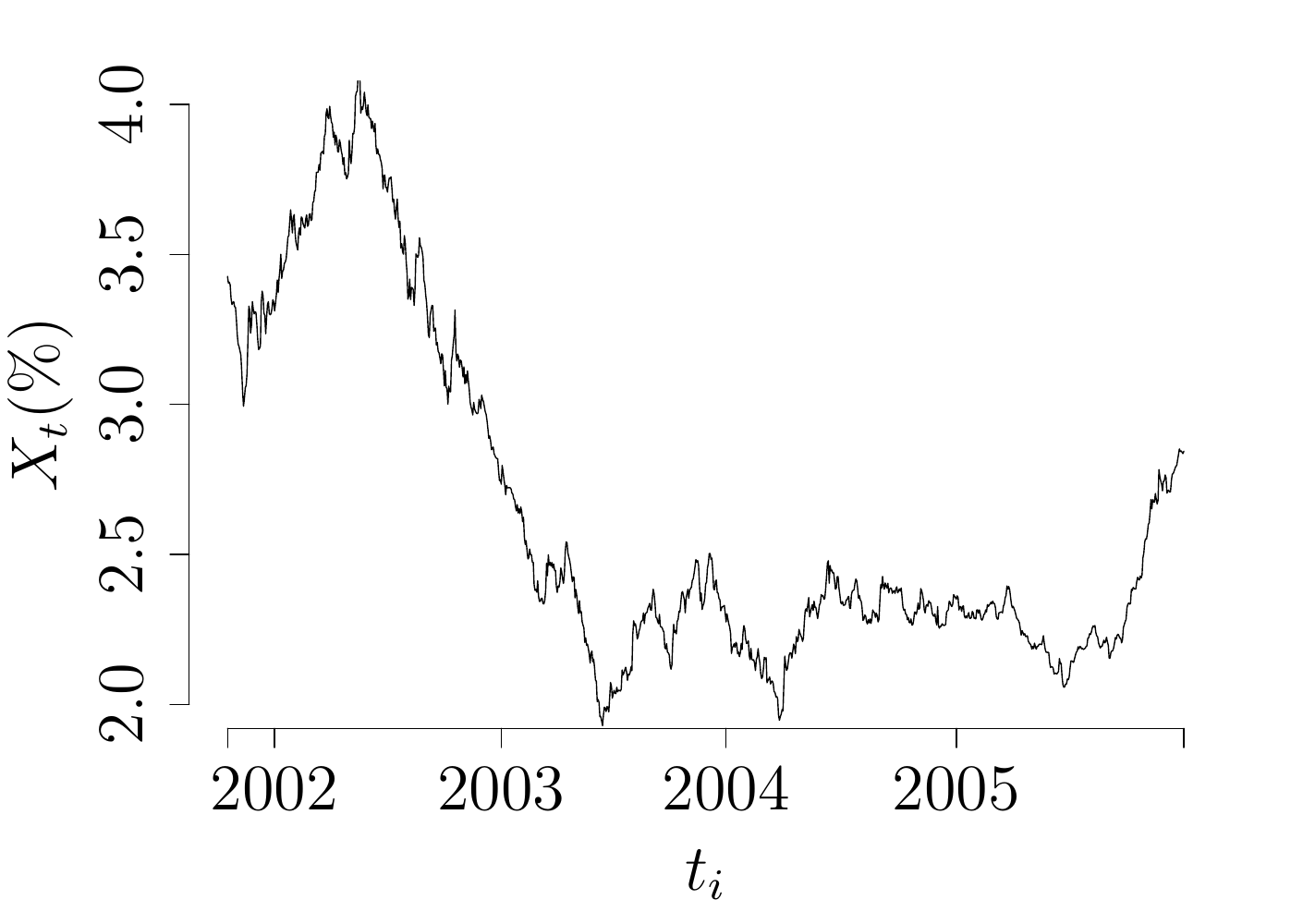}
		\captionsetup{font=footnotesize,textfont=it}
		\caption{Euribor 12 months.}
		\label{fig:euribor12m}
	\end{subfigure}
	\caption{Daily evolution of Euribor series for the time period between October 15\textsuperscript{th} 2001 and December 30\textsuperscript{th} 2005. Sample size for each dataset is $n=1\,077$.}
	\label{fig:euribor}
\end{figure}

In this section, we consider the Euribor (Euro Interbank Offered Rate) interest rate series corresponding to four maturities (three, six, nine and twelve months), see Figure~\ref{fig:euribor}. The four datasets expand from October 15\textsuperscript{th} 2001 to December 30\textsuperscript{th} 2005 (sample size of $n = 1\,077$). We fit the CKLS with stochastic volatility in~\eqref{eq:SDE_CKLS_dis}, as different models can be nested within this unrestricted model, and test the goodness of fit of the model in terms of the parametric form of the volatility function. 

Table~\ref{tab:estEuribor} shows the parameter estimations for the CKLS-OU model, with the associated standard error in parentheses. The values verify the trait usually associated with interest rate time series, that is, persistence, both for the interest rate $r_t$ and volatility $\sigma_t$ equations. Regarding the parameter $\gamma$, which controls the relationship between the interest rate level and the volatility, for all maturities we have $\gamma > 1$. This indicates that the volatility tends to increase as the interest rate $r_t$ raises.

Regarding the goodness-of-fit test, Table~\ref{tab:EURpval} shows the $p$-values for the parametric form of the volatility function, both for the Kolmogorov-Smirnov and Cramér-von Mises statistics, which exhibit minor discrepancies. In \cite{lopez2021parametric} the same datasets were used to fit a CKLS model with deterministic volatility function and the null hypothesis for the parametric form of the diffusion function was strongly rejected for the four maturities. However, when considering a more flexible diffusion function with stochastic volatility, we do not reject the null hypothesis, suggesting that a model that incorporates stochastic volatility may adequately explain the dynamics of the series. Rejecting a deterministic volatility function in favor of a stochastic function indicates that the volatility evolution is not exclusively tied to the level of the short rate, but rather the process is governed by dynamic factors. This was discussed in the financial literature (see \citealp{ait1996interest}; \citealp{brenner1996another}; \citealp{andersen1997estimating}; \citealp{koedijk1997dynamics}; \citealp{gallant1998reprojecting}), where less restrictive models were proposed, including the two-factor model or even multi-factor models of the short rate \citep{andersen1997stochastic}.

\begin{table}[H]
	\centering
	\begin{tabular}{ccccccc}
		\toprule
		Parameters: & $\alpha$ & $\beta$ & $\gamma$ & $\phi_0$ & $\phi_1$ & $\xi$ \\
		\midrule
		3 months  & $1.6546$   & $0.7733$   & $1.6483$   & $-0.2821$  & $0.9674$   & $7.2056$   \\
		          & $(0.5980)$ & $(0.2631)$ & $(0.6711)$ & $(0.1696)$ & $(0.0177)$ & $(1.7736)$ \\
		6 months  & $1.5139$   & $0.6813$   & $1.4219$   & $-0.0812$  & $0.9878$   & $2.4606$   \\
		          & $(0.7567)$ & $(0.3214)$ & $(0.5848)$ & $(0.0728)$ & $(0.0103)$ & $(0.8608)$ \\
		9 months  & $1.6850$   & $0.7218$   & $1.7627$   & $-0.0494$  & $0.9919$   & $1.3615$   \\
		          & $(1.0196)$ & $(0.4218)$ & $(0.4122)$ & $(0.0479)$ & $(0.0079)$ & $(0.5332)$ \\
		12 months & $1.9218$   & $0.7868$   & $1.6729$   & $-0.0344$  & $0.9937$   & $1.1163$   \\
		          & $(1.2384)$ & $(0.4977)$ & $(0.3914)$ & $(0.0358)$ & $(0.0066)$ & $(0.4468)$ \\
		\bottomrule
	\end{tabular}%
	\caption{\label{tab:estEuribor} Parameter estimates and standard errors (in parentheses) for the CKLS process with stochastic volatility, fitted to Euribor series.}%
\end{table}%

\begin{table}[H]
	\centering
	\begin{tabular}{lcccc}
		\toprule
		Maturity & 3 months & 6 months & 9 months & 12 months \\
		\midrule
		p-value Kolmogorov-Smirnov   & $0.464$ & $0.266$ & $0.319$ & $0.189$  \\       
		p-value Cramér-von Mises     & $0.550$ & $0.215$ & $0.422$ & $0.113$  \\       
		\bottomrule
	\end{tabular}
	\caption{\label{tab:EURpval} $p$-values for the goodness-of-fit test for the CKLS-OU parametric form of the diffusion function.
	}
\end{table}

\section{Conclusions} \label{sec:6}

We reviewed parametric estimation methods for two-factor continuous-time stochastic volatility models. The continuous time nature of the process does complicate the parameter estimation, as available data are registered in discrete time points. As a consequence, parameters are subject to discretization bias and this, combined with the presence of a latent factor, challenges estimation. We discussed a comparative study of three estimation methods --namely, MCMC, Kalman Filter and particle filter-- under different settings. The close performance of the procedures, together with the computationally demanding and model-dependent implementation of simulation methods, makes the Kalman Filter a computational efficient estimation method to use in goodness-of-fit testing procedures. Furthermore, the space-state model structure allows to easily implement a bootstrap procedure.
We proposed goodness-of-fit tests for the parametric form of the drift and volatility functions, based on a residual marked empirical process. The tests showed great power through several alternative hypotheses and was well calibrated under null hypotheses and its implementation, regarding the computation of the test statistic and the bootstrap resampling scheme, is quite straightforward.
The application to real data demonstrated that the incorporation of stochastic volatility to diffusion models does capture the features of interest rate time series, unlike deterministic volatility functions, suggesting that the volatility depends on an additional factor that varies independently of the short rate level.

\bibliography{biblio}
\bibliographystyle{apalike} 

\end{document}